\documentclass[12pt]{article}
\usepackage{graphicx,rotating,epsfig,amsmath,amssymb,color,multirow,pstricks,pst-plot,comment}
\newlength{\digitwidth} 

\newcommand{\jpsi}{J/$\psi$}
\newcommand{\psip}{$\psi^\prime$}
\newcommand{\chicAll}{$\chi_c$}

\newcommand{\chicOne}{$\chi_{c1}$}
\newcommand{\chicTwo}{$\chi_{c2}$}
\newcommand{\upsAll}{$\Upsilon$}
\newcommand{\upsOneS}{$\Upsilon(1S)$}
\newcommand{\upsTwoS}{$\Upsilon(2S)$}
\newcommand{\upsThreeS}{$\Upsilon(3S)$}
\newcommand{\chibAll}{$\chi_b$}

\newcommand{\pt}{$p_{\rm T}$}

\newcommand{\ccbar}{$c \bar{c}$}
\newcommand{\bbbar}{$b \bar{b}$}
\newcommand{\qqbar}{$Q \bar{Q}$}

\newcommand{\cinst}[2]{$^{\mathrm{#1)}}$~#2\par}
\newcommand{\crefi}[1]{$^{\mathrm{#1)}}$}
\newcommand{\HRule}{\rule{0.4\linewidth}{0.3mm}}

\begin{document}

%%%%%%%%%%%%%%%%%%%%%%%%% CERN PREPRINT COVER PAGE
\begingroup
\thispagestyle{empty} \baselineskip=14pt
\parskip 0pt plus 5pt

\begin{center}
{\large EUROPEAN LABORATORY FOR PARTICLE PHYSICS}
\end{center}

\bigskip
\begin{flushright}
CERN--PH--EP\,/\,2010--0xx\\
June 14, 2010
\end{flushright}

\bigskip
\begin{center}
{\Large\bf \boldmath
Towards the experimental clarification\\[3mm]
of quarkonium polarization
}

\bigskip\bigskip

Pietro Faccioli\crefi{1},
Carlos Louren\c{c}o\crefi{2},
Jo\~ao Seixas\crefi{1,3}
and Hermine K.\ W\"ohri\crefi{1}

\bigskip\bigskip\bigskip
\textbf{Abstract}

\end{center}

\begingroup
\leftskip=0.4cm \rightskip=0.4cm \parindent=0.pt
%%%%
%%%\title{Towards the experimental clarification of quarkonium
%%%  polarization}
%%%%
%%%\author{Pietro Faccioli\inst{1} \and Carlos Louren\c{c}o\inst{2}
%%%\and Jo\~ao Seixas\inst{1,3} \and Hermine K.\ W\"ohri\inst{1}
%%%}
%%%%
%%%\institute{Laborat\'orio de Instrumenta\c{c}\~ao e F\'{\i}sica Experimental de
%%%  Part\'{\i}culas (LIP), Lisbon, Portugal
%%% \and 
%%%CERN, Geneva, Switzerland
%%%\and
%%%Physics Department, Instituto Superior T\'ecnico (IST),
%%%  Lisbon, Portugal
%%%}
%%%%
%%%\date{Received: June 14, 2010 / Revised version: date}
%%%%
%%%\abstract{

  We highlight issues which are often underestimated in the
  experimental analyses on quarkonium polarization: the relation
  between the parameters of the angular distributions and the angular
  momentum composition of the quarkonium, the importance of the choice
  of the reference frame, the interplay between observed decay and
  production kinematics, and the consequent influence of the
  experimental acceptance on the comparison between experimental
  measurements and theoretical calculations. Given the puzzles
  raised by the available experimental results, new measurements must
  provide more detailed information, such that physical conclusions
  can be derived without relying on model-dependent assumptions. We
  describe a frame-invariant formalism which minimizes the dependence
  of the measurements on the experimental acceptance, facilitates the
  comparison with theoretical calculations, and probes systematic
  effects due to experimental biases. This formalism is a direct and
  generic consequence of the rotational invariance of the dilepton
  decay distribution and is independent of any assumptions specific to
  particular models of quarkonium production. The use of this improved
  approach, which exploits the intrinsic multidimensionality of the
  problem, will significantly contribute to a faster progress in our
  understanding of quarkonium production, especially if adopted as a
  common analysis framework by the LHC experiments, which will soon
  perform analyses of quarkonium polarization in proton-proton
  collisions.
%
%%%\PACS{
%%%      {11.80.Cr}{Kinematical properties (helicity and invariant
%%%        amplitudes, kinematic singularities, etc.)}   \and
%%%      {12.38.Qk}{Experimental tests of QCD}   \and
%%%      {13.20.Gd}{Decays of \jpsi, $\Upsilon$, and other quarkonia}   \and
%%%      {13.85.Qk}{Inclusive production with identified leptons,
%%%        photons, or other nonhadronic particles}
%%%     } % end of PACS codes
%%%} %end of abstract
\bigskip
%%%\maketitle
%
%
%
\endgroup
\vspace{1cm}
\begin{center}
\emph{Submitted to Euro. Phys. J. C}
\end{center}
\vfill
\begin{flushleft}
\HRule\\

\cinst{1} {Laborat\'orio de Instrumenta\c{c}\~ao e F\'{\i}sica Experimental de
  Part\'{\i}culas (LIP),\\ ~~~Lisbon, Portugal}
\cinst{2} {CERN, Geneva, Switzerland}
\cinst{3} {Physics Department, Instituto Superior T\'ecnico (IST),
  Lisbon, Portugal}
\end{flushleft}
\endgroup

\newpage

\sloppy

%%%%%%%%%%%%%%%%%%%%%%%%%%%%%%%%
\section{Introduction}
\label{sec:intro}

Detailed studies of quarkonium prodution should provide significant
progress in our understanding of quantum chromodynamics
(QCD)~\cite{bib:YellowRep-QWG}.  However, our present understanding of
this physics topic is rather limited, despite the multitude of
experimental data accumulated over more than 30 years.
The \pt\ differential \jpsi\ and \psip\ direct
production cross sections measured (in the mid 1990's) by CDF, in
${\rm p}\bar{\rm p}$ collisions at 1.8~TeV~\cite{bib:cdf1-psis}, were
seen to be around 50 times larger than the available expectations,
based on leading order calculations made in the scope of the Colour
Singlet Model.  The non-relativistic QCD (NRQCD)
framework~\cite{bib:NRQCD}, where quarkonia can also be produced as
\emph{coloured} quark pairs, succeeded in describing the measurements,
opening a new chapter in the studies of quarkonium production physics.
However, these calculations depend on non-perturbative parameters, the
long distance colour octet matrix elements, which have been freely
adjusted to the data, thereby decreasing the impact of the resulting
agreement between data and calculations.
More recently, calculations of next-to-leading-order (NLO) QCD
corrections to colour-singlet quarkonium production showed an
important increase of the high-\pt\ rate, significantly decreasing the
colour-octet component needed to reproduce the quarkonium production
cross sections measured at the Tevatron~\cite{bib:lansberg-HP08}.

Given this situation, differential cross sections are clearly
insufficient information to ensure further progress in our
understanding of quarkonium production. Experimental studies of the
polarization of the $J^{PC}=1^{--}$ quarkonium states, which decay
into lepton pairs, will certainly provide very useful complementary
information. In fact, the competing mechanisms dominating in the
different theoretical approaches lead to very different expected
polarizations of the produced quarkonia. On one hand, the NRQCD
calculations~\cite{bib:BK,bib:Lei,bib:BKL}, dominated by the
colour-octet component, predict that, at Tevatron or LHC energies and
at asymptotically high \pt, the directly produced \psip\ and \jpsi\
mesons are produced almost fully \emph{transversely} polarized (i.e.\
with dominant angular momentum component $J_z = \pm 1$) with respect
to their own momentum direction (the \emph{helicity frame}). On the
other hand, according to the new NLO calculations of colour-singlet
quarkonium production~\cite{bib:lansberg-HP08} these states should
show a strong \emph{longitudinal} ($J_z = 0$) polarization component.

Having two very different theoretical predictions appears to be an
ideal situation when seen from an experimentalist's perspective, as
one may think that it should be relatively straightforward to
discriminate between the two theory frameworks using experimental
measurements.  Somewhat surprisingly, however, this is not the case.
In fact, the present experimental knowledge is incomplete and
contradictory.
Studies of the \psip\ polarization have been published on the basis of
data collected by the CDF~II experiment~\cite{bib:CDFpol2}.  Unfortunately,
the large experimental uncertainties caused by the small size of the
data samples prevent from drawing meaningful conclusions.
In principle, a more precise test of the theoretical predictions
should be provided by the \jpsi\ data, given their much higher
statistical accuracy. However, the experimental perspective is more
complicated in this case, because a significant fraction (around one
third~\cite{bib:feeddown}) of promptly produced \jpsi\ mesons (i.e.\
excluding contributions from B hadron decays) comes from \chicAll\ and \psip\
feed-down decays. This sizeable source of indirectly produced \jpsi\
mesons is not subtracted from the current measurements, and its
kinematic dependence is not precisely known.
Despite this limitation, it seems safe to say that the pattern
measured by CDF~\cite{bib:CDFpol2} of a slightly longitudinal
polarization of the inclusive prompt \jpsi's is incompatible with any
of the two theory approaches mentioned above.
The situation is further complicated by the intriguing lack of
continuity between fixed-target and collider results, which can only
be interpreted in the framework of some specific (and speculative)
assumptions still to be tested~\cite{bib:pol}.

The \bbbar\ system should satisfy the non-relativistic approximation
much better than the \ccbar\ case. For this reason, the \upsAll\ data
are expected to represent the most decisive test of NRQCD. However,
the comparison with the existing \upsOneS\ polarization data from
Tevatron~\cite{bib:upsCDF,bib:upsD0} is far from conclusive. The
results indicate that, for $p_{\rm T} < 15$~GeV$/c$, the \upsOneS\ is
produced either unpolarized (CDF) or longitudinally polarized (D0) in
the helicity frame, and this discrepancy cannot be reasonably
attributed to the different rapidity windows covered by the two
experiments. Furthermore, the precision of the data for \pt\ higher
than $15$~GeV$/c$ is not sufficient to provide a significant test of
the crucial hypothesis that very high-\pt\ quarkonia, produced by the
fragmentation of an outgoing (almost on-shell) gluon, are fully
transversely polarized along their own direction.
At lower energy and \pt, the E866 experiment~\cite{bib:e866_Ups} has
shown yet a different polarization pattern: the \upsTwoS\ and
\upsThreeS\ states have \emph{maximal} transverse polarization, with
no significant dependence on transverse or longitudinal momentum,
\emph{with respect to the direction of motion of the colliding
  hadrons} (Collins-Soper frame). Unexpectedly, the \upsOneS, whose
spin and angular momentum properties are identical to the ones of the
heavier \upsAll\ states, is, instead, found to be only weakly
polarized. These results give interesting physical indications.
First, the maximal polarization of \upsTwoS\ and \upsThreeS\ along the
direction of the interacting particles places strong constraints on
the topology and spin properties of the underlying elementary
production process. Second, the small \upsOneS\ polarization suggests
that the bottomonium family may have a peculiar feed-down hierarchy,
with a very significant fraction of the lower mass state being
produced indirectly; at the same time, the polarization of the
\upsAll's coming from \chibAll\ decays should be substantially
different from the polarization of the directly produced ones.

This rather confusing situation demands a significant improvement in
the accuracy and detail of the polarization measurements, ideally
distinguishing between the properties of directly and indirectly
produced states.
We remind that the lack of a consistent description of the
polarization properties represents today's biggest uncertainty in the
simulation of the LHC quarkonium production measurements and will
probably be the largest contribution to the systematic error affecting
the measurements of quarkonium production cross sections and kinematic
distributions.  Indeed, the probability that the detector accepts
lepton pairs resulting from decays of quarkonium states is strongly
dependent on the polarization of those states.  Therefore, even from a
purely experimental point of view it is very unsatisfactory that
essential properties of these objects, such as kinematic details of
how they decay into lepton pairs (on which their reconstruction is
based), are subjected to such a high degree of uncertainty.

It is true that measurements of the quarkonium decay angular
distributions are challenging, multi-dimensional kinematic problems,
which require large event samples and a very high level of
accuracy in the subtraction of spurious kinematic correlations induced
by the detector acceptance. The complexity of the experimental
problems which have to be faced in the polarization measurements is
testified, for example, by the disagreement between the CDF results
obtained in Run~I and Run~II for the
\jpsi~\cite{bib:CDFpol1,bib:CDFpol2} and by the contradictory results
obtained by CDF and D0 for the \upsOneS.
However, it is also true, as we shall emphasize in this paper,
that most experiments have exploited, and presented in the published
reports, only a fraction of the physical information derivable from
the data. This happens, for example, when the measurement is performed
in only one polarization frame and is limited to the polar projection
of the decay angular distribution. As we have already argued in
Ref.~\cite{bib:pol}, these incomplete measurements do not allow
definite physical conclusions.  At best, they confine such conclusions
to a genuinely model-dependent framework. Moreover, such a fragmentary
description of the observed physical process obviously reduces the
chances of detecting possible biases induced by not fully controlled
systematic effects.

In this paper we review the mathematical framework for the description
of the observable polarization of quarkonium states decaying
electromagnetically into lepton pairs.
We focus our attention on aspects that need to be taken in
consideration in the analysis of the data, so as to maximize the
physical significance of the measurement and provide all elements for
its unambiguous interpretation within any theoretical framework.
By increasing the level of detail of the physical information
extracted from the data, the proposed methodologies also offer the
possibility of performing consistency checks which can expose
unaccounted detector or analysis biases.
The only relevant theoretical ingredients of our discussion are the
quantum properties of angular momentum and basic conservation rules of
the electromagnetic interaction (parity, fermion chirality). All the
results presented here are, therefore, valid in general for any
quarkonium production mechanism.

In Section~\ref{sec:concepts} we define the concept of polarization
and give simple examples of how basic production mechanisms can lead
to the formation of polarized quarkonium states. We then focus on the
dilepton decay distribution of $^3S_1$ quarkonia, a relatively simple
case, and provide detailed geometric and kinematic considerations.
In Section~\ref{sec:distribution} we recall the basic principles
leading to the general expression describing the angular distribution
of the decay products, while in Section~\ref{sec:frame_dependence} we
describe how the observed anisotropy parameters depend on the choice
of the reference frame.
Section~\ref{sec:kinematics} is devoted to a detailed description of
how the production kinematics influences the observed polarization,
depending on the quarkonium momentum and on the observation frame.  We
also discuss quantitatively the influence of the intrinsic parton
transverse momenta on the polarization measurement when the natural
axis is along the relative flight direction of the colliding partons.
In Section~\ref{sec:invariant} we derive the existence of a
frame-independent identity which relates the observable parameters of
the decay distribution to one frame-invariant polarization parameter.
We discuss how this relation, formally including the Lam-Tung
identity~\cite{bib:LamTung} as a particular case, improves the
representation of polarization results and can be used to perform
consistency checks in the experimental analyses.
We continue with some remarks, given in Section~\ref{sec:kt}, on how
the existence of intrinsic parton transverse momentum affects the
polarization measurement.
We conclude, in Section~\ref{sec:examples}, with a few examples
inspired from existing experimental measurements, which should provide
concrete evidence for the usefulness of the approaches discussed in
this paper, in view of ensuring an improved understanding of
quarkonium production.

%%%%%%%%%%%%%%%%%%%%%%%%%%%%%%%%%%%%%%%%%%%%%%%%
\section{Basic polarization concepts}
\label{sec:concepts}

\begin{figure*}[htb]
\centering
\resizebox{1.0\linewidth}{!}{%
\includegraphics{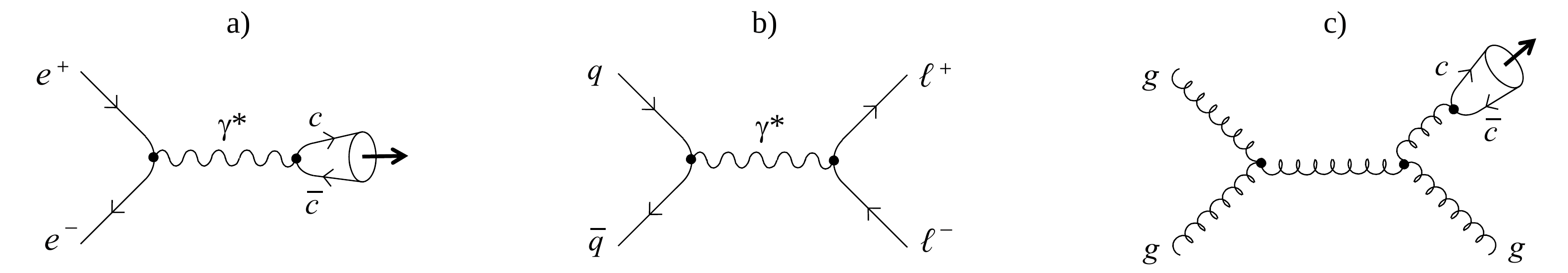}}
\caption{Examples of leading-order diagrams for
  production mechanisms giving rise to observable polarizations:
  (a)~vector quarkonium production in electron-positron annihilation;
  (b)~Drell-Yan production in quark-antiquark annihilation;
  (c)~quarkonium production by gluon fragmentation to colour-octet
  \ccbar.}
\label{fig:processes}
\end{figure*}

Because of angular momentum conservation and basic symmetries of the
electromagnetic and strong interactions, a particle produced in a
certain superposition of elementary mechanisms may be observed
preferentially in a state belonging to a definite subset of the
possible eigenstates of the angular momentum component $J_z$ along a
characteristic quantization axis. When this happens, the particle is
said to be polarized. Figure~\ref{fig:processes} shows examples of
leading-order diagrams of elementary production processes giving rise
to different types of polarizations.

%%%%%%%%%%%%%%
%%%%%%%%%%%%%%

Vector ($J=1^{--}$) quarkonia have the same charge-parity as an
electron-positron pair and can be produced in electron-positron
annihilation, via an intermediate photon (Fig.~\ref{fig:processes}\,a).
The states originating from this process are polarized, as a
consequence of helicity conservation, a general property of QED in the relativistic
(massless) limit. The dynamics of the coupling of electrons to
photons is, in fact, described by terms of the form $\overline{u}
\gamma^\mu u = \overline{u}_\mathrm{L} \gamma^\mu u_\mathrm{L} +
\overline{u}_\mathrm{R} \gamma^\mu u_\mathrm{R}$, where $\gamma^\mu$
are the Dirac matrices, $u$ is the electron spinor, and L (R) indicate
its left-handed (right-handed) chiral components. Terms with opposite
chiral components are absent, meaning that the fermion chirality is
preserved in the interaction with a photon. When the fermions are
assumed to have zero mass, so that the direction of their momenta
cannot be reversed by any Lorentz transformation, left-handed and
right-handed chiral components become eigenstates of the helicity
operator $h = \vec{S} \cdot \vec{p} / |\vec{p}|$, 
corresponding to the projection of the
spin on the momentum direction.  In this case, chirality conservation
becomes helicity conservation. In the diagram of
Fig.~\ref{fig:processes}\,a, this rule implies that the annihilating
electron and positron must have opposite helicities, because the
intermediate photon has zero (fermion) helicity. Since in the
laboratory their momenta are opposite, their spins must be parallel.
Because of angular momentum conservation, the produced quarkonium has,
thus, angular momentum component $J_z = \pm 1$ along the direction of
the colliding leptons. This precise QED prediction (the relative
amplitude for the $J_z = 0$ component is of order $m_e/E_e \simeq 3
\times 10^{-4}$ for \jpsi\ production and smaller for \upsAll\
production) is commonly used as a base assumption in quarkonium
measurements in electron-positron annihilations (as, for example, in
the recent analysis of Ref.~\cite{bib:chic_angulardistr_CLEO}). The
fact that the dilepton system coupled to a photon is a pure $J_z = \pm
1$ state is also an essential ingredient in the determination of the
expression for the dilepton-decay angular distributions of vector
quarkonia (see Section~\ref{sec:distribution}).

The same reasoning can be applied to the production of Drell-Yan
lepton pairs in quark-antiquark annihilation
(Fig.~\ref{fig:processes}\,b): the quark and antiquark, in the limit of
vanishing masses, must annihilate with opposite helicities, resulting
in a dilepton state having $J_z = \pm 1$ along the direction of their
relative velocity. The experimental verification of this basic
mechanism has reached an impressive level of
accuracy~\cite{bib:e866_Ups}.  Quark helicity is conserved also in
QCD, when the masses can be neglected. Similarly to the Drell-Yan
case, quarkonia originating from quark-antiquark annihilation
(into intermediate gluons) will thus tend, provided they are produced
alone, to have their angular momentum vectors ``aligned'' ($J_z = \pm
1$) along the beam direction. This prediction is in good agreement
with
%%% well compatible with
the \chicOne, \chicTwo\ and \psip\ polarizations measured in
low-energy proton-antiproton
collisions~\cite{bib:chic_angulardistr_Fermilab,bib:psip_E835}.

At very high \pt, quarkonium production at hadron colliders should
mainly proceed by gluon fragmentation~\cite{bib:gluonFragm}. In NRQCD,
heavy-quark velocity scaling rules for the non perturbative matrix
elements, combined with the $\alpha_S$ and $1/p_{\rm T}$ power
counting rules for the parton cross sections, predict that \jpsi\ and
\psip\ production at high \pt\ is dominated by gluon fragmentation
into the color-octet state $c\bar{c}[^3S^{(8)}_1]$
(Fig.~\ref{fig:processes}\,c).  Transitions of the gluon to other
allowed colour and angular momentum configurations, containing the
\ccbar\ in either a colour-singlet or a colour-octet state, with spin
$S = 0, 1$ and angular momentum $L = 0, 1, 2, \ldots$, as well as
additional gluons ($c\bar{c}[^1S^{(8)}_0]g$, $c\bar{c}[^3P^{(8)}_J]g$,
$c\bar{c}[^3S^{(1)}_1]gg$, etc.), are more and more suppressed with
increasing \pt. Up to small corrections, the fragmenting gluon is
believed to be on shell and have, therefore, helicity $\pm 1$. This
property is inherited by the $c\bar{c}[^3S^{(8)}_1]$ state and remains
intact during the non-perturbative transition to
the colour-neutral physical state, via soft-gluon emission.
In this model, the observed charmonium has, thus, angular momentum
component $J_z = \pm 1$, this time not along the direction of the
beam, but along its own flight direction.

``Unpolarized'' quarkonium has the same probability, $1/(2J+1)$, to be
found in each of the angular momentum eigenstates, $J_z = -J, -J+1,
\ldots, +J$. This is the case, for example, in the colour evaporation
model~\cite{bib:colour_evap}. In this framework, similarly to NRQCD,
the \qqbar\ pair is produced at short distances in any colour and
angular momentum configuration. However, contrary to NRQCD, no
hierarchy constraints are imposed on these configurations, so that the
cross section turns out to be dominated by \qqbar\ pairs with
vanishing angular momentum ($^1S_0$), in either colour-singlet or
colour-octet states. In their long distance evolution through soft
gluon emissions, $J=0$ states get their colour randomized, assuming
the correct quantum numbers of the physical quarkonium.  As a result,
the final angular momentum vector $\vec{J}$ has no preferred
alignment.

In two-body decays (such as the $^3S_1 \rightarrow \ell^+ \ell^-$ case
considered in this paper), the geometrical shape of the angular
distribution of the two decay products (emitted back-to-back in the
quarkonium rest frame) reflects the polarization of the quarkonium
state.  A spherically symmetric distribution would mean that the
quarkonium would be, on average, unpolarized.  Anisotropic
distributions signal polarized production.

%
%%%%%%%%%%%%%%%%%%%%%%%%%%%%%%%%%%%%%%%%%%%%%%%%%%%%%%%%%%%%%%%%%%%%%%%%%%%%%%%%%%%%%%%
\begin{figure}[htb]
\centering
\resizebox{0.5\linewidth}{!}{%
\includegraphics{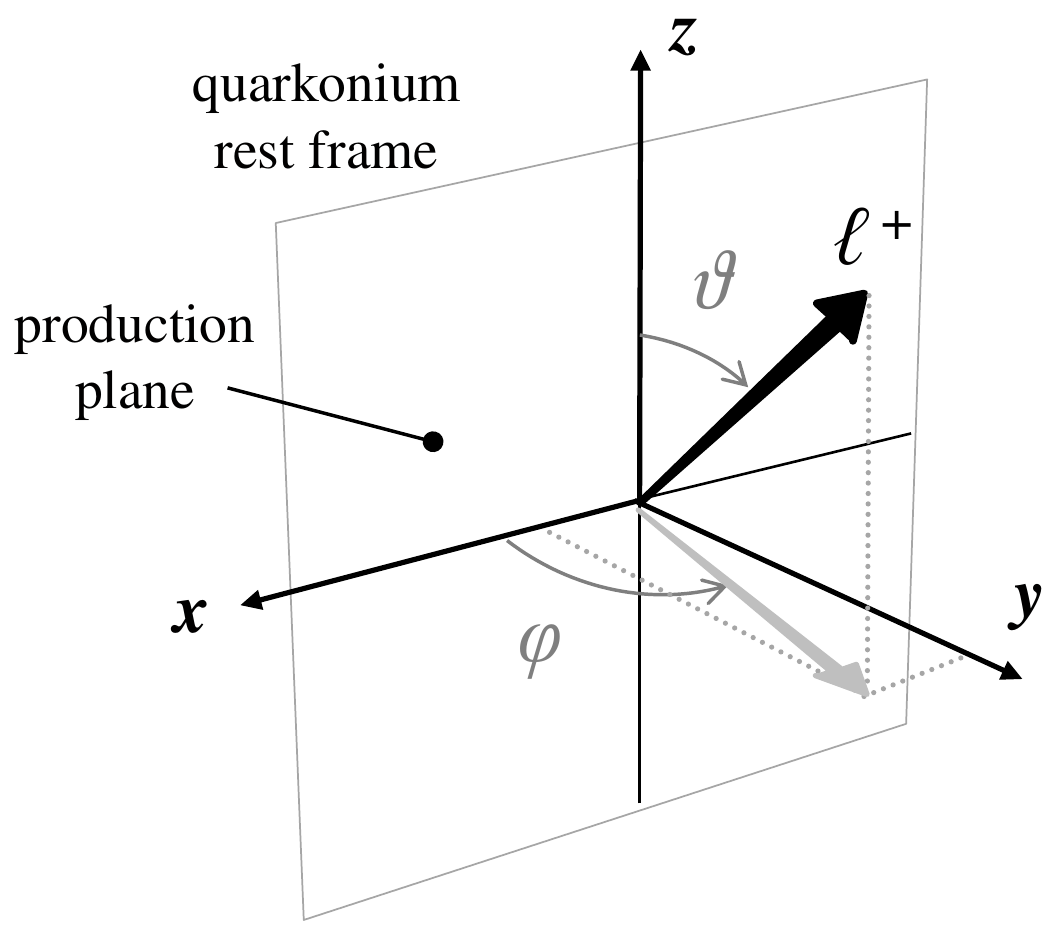}}
\caption{\label{fig:coordinates} The coordinate system for the measurement of a
two-body decay angular distribution in the quarkonium rest frame. The $y$ axis
is perpendicular to the plane containing the momenta of the colliding beams.
The polarization axis $z$ is chosen according to one of the possible
conventions described in Fig.~\ref{fig:frames}.}
\end{figure}
%%%%%%%%%%%%%%%%%%%%%%%%%%%%%%%%%%%%%%%%%%%%%%%%%%%%%%%%%%%%%%%%%%%%%%%%%%%%%%%%%%%%%%%
%
The measurement of the distribution requires the choice of a
coordinate system, with respect to which the momentum of one of the
two decay products is expressed in spherical coordinates.  In
inclusive quarkonium measurements, the axes of the coordinate system
are fixed with respect to the physical reference provided by the
directions of the two colliding beams as seen from the quarkonium rest
frame.  Figure~\ref{fig:coordinates} illustrates the definitions of
the polar angle $\vartheta$, determined by the direction of one of the
two decay products (e.g.\ the positive lepton) with respect to the
chosen polar axis, and of the azimuthal angle $\varphi$, measured with
respect to the plane containing the momenta of the colliding beams
(``production plane'').
%
%%%%%%%%%%%%%%%%%%%%%%%%%%%%%%%%%%%%%%%%%%%%%%%%%%%%%%%%%%%%%%%%%%%%%%%%%%%%%%%%%%%%%%%
\begin{figure*}[htb]
\centering
\resizebox{0.8\linewidth}{!}{%
\includegraphics{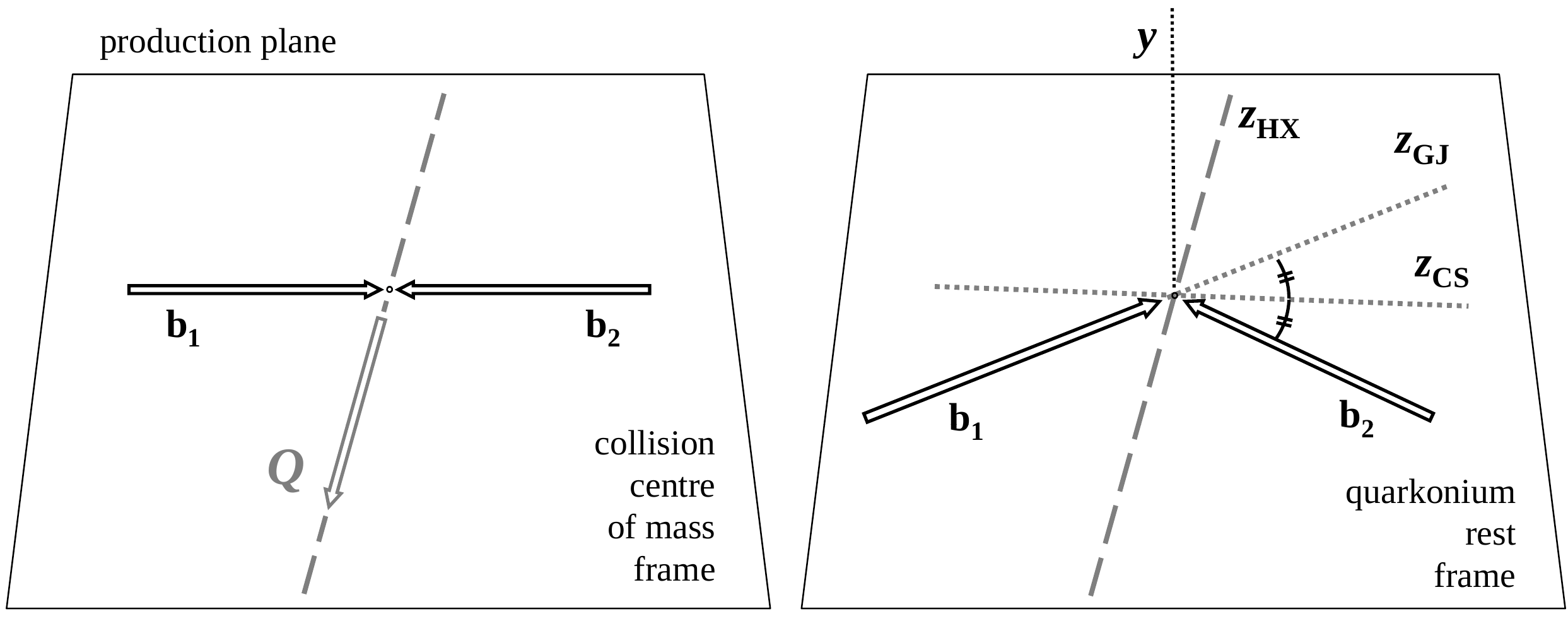}}
\caption{\label{fig:frames} Illustration of three different definitions of the
polarization axis $z$ (CS: Collins-Soper, GJ: Gottfried-Jackson, HX: helicity)
with respect to the directions of motion of the colliding beams
($\mathrm{b}_1$, $\mathrm{b}_2$) and of the quarkonium ($\mathcal{Q}$).}
\end{figure*}
%%%%%%%%%%%%%%%%%%%%%%%%%%%%%%%%%%%%%%%%%%%%%%%%%%%%%%%%%%%%%%%%%%%%%%%%%%%%%%%%%%%%%%%
%
The actual definition of the decay reference frame with respect to the
beam directions is not unique.  Measurements of the quarkonium decay
distributions have used three different conventions for the
orientation of the polar axis (see Fig.~\ref{fig:frames}): the
direction of the momentum of one of the two colliding beams
(Gottfried-Jackson frame~\cite{bib:gott_jack}, GJ), the opposite of
the direction of motion of the interaction point (i.e.\ the flight
direction of the quarkonium itself in the center-of-mass of the
colliding beams: helicity frame, HX) and the bisector of the angle
between one beam and the opposite of the other beam (Collins-Soper
frame~\cite{bib:coll_sop}, CS). The motivation of this latter
definition is that, in hadronic collisions, it coincides with the
direction of the relative motion of the colliding partons, when their
transverse momenta are neglected (the validity and limits of this
approximation are discussed in detail in Section~\ref{sec:kt}).
For our considerations, we will take the HX and CS frames as two
extreme (physically relevant) cases, given that the GJ polar axis
represents an intermediate situation.
We note that these two frames differ by a rotation
of $90^\circ$ around the $y$ axis when the quarkonium is produced at high \pt\
and negligible longitudinal momentum ($p_{\rm T} \gg |p_{\rm
  L}|$). All definitions become coincident in the limit of zero
quarkonium \pt. In this limit, moreover, for symmetry reasons any
azimuthal dependence of the decay distribution is physically
forbidden.

We conclude this section by defining the somewhat misleading
nomenclature which is commonly used (and adopted, for convenience,
also in this paper) for the polarization of vector mesons. These
particles share the quantum numbers of the photon and are therefore
said, by analogy with the photon, to be ``transversely'' polarized
when they have spin projection $J_z = \pm 1$. The counterintuitive
adjective originally refers to the fact that the electromagnetic field
carried by the photon oscillates in the transverse plane with respect
to the photon momentum, while the photon \emph{spin} is aligned
\emph{along} the momentum. ``Longitudinal'' polarization means $J_z =
0$. By further extension, the same terms are also used to describe the
``spin alignment'' of vector quarkonia not only with respect to their
own momenta (HX frame), but also with respect to any other chosen
reference direction (such as the GJ or CS axes).

%%%%%%%%%%%%%%%%%%%%%%%%%%%%%%%%%%%%%%%%%%%%%%%%%%%
%%%%%%%%%%%%%%%%%%%%%%%%%%%%%%%%%%%%%%%%%%%%%%%%%%%
\section{Dilepton decay angular distribution}
\label{sec:distribution}

Vector quarkonia, such as the \jpsi, \psip\ and $\Upsilon(nS)$ states,
can decay electromagnetically into two leptons. The reconstruction of
this channel represents the cleanest way, both from the experimental
and theoretical perspectives, of measuring their production yields and
polarizations. In this and the following sections we discuss how to
determine experimentally the ``spin alignment'' of a vector quarkonium
by measuring the dilepton decay angular distribution.
%%% in a chosen observation frame.
For convenience we mention explicitly the \jpsi\ as the decaying
particle, but considerations and results are valid for any $J=1^{--}$
state.

\begin{figure}[htb]
\centering
\resizebox{0.5\linewidth}{!}{%
\includegraphics{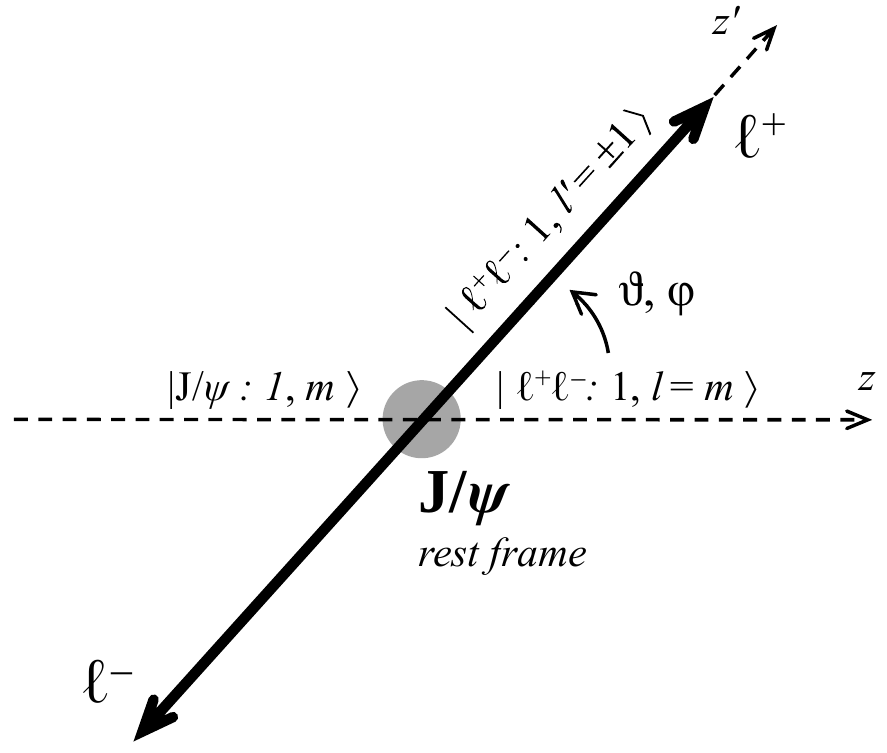}}
\caption{\label{fig:psidecay} Sketch of the decay $\mathrm{J}/\psi
  \rightarrow \ell^+ \ell^-$, showing the notations used in the text
  for axes, angles and angular momentum states.}
\end{figure}

We begin by studying the case of a single production ``subprocess'',
here defined as a process where the \jpsi\ is formed as a given
superposition of the three $J = 1$ eigenstates, $J_z = +1, -1, 0$
with respect to the polarization axis~$z$:
\begin{equation}
  | V \rangle =  b_{+1} \, |+1\rangle + b_{-1} \, |-1\rangle + b_{0} \, |0\rangle \,
  . \label{eq:state}
\end{equation}
The calculations are performed in the \jpsi\ rest frame, where the
common direction of the two leptons define the reference axis
$z^\prime$, oriented conventionally along the direction of the
positive lepton. The adopted notations for axes, angles and angular
momentum states are illustrated in Fig.~\ref{fig:psidecay}.
%
%%%%%%%%%%%%%%%%%%%%%%%%%%%%%%%%%%%%%%%%%%%%%%%%%%%%%%%%%%%%%%%%%%%%%%%%%%%%%%%%%%%%%%%
%%%%%%%%%%%%%%%%%%%%%%%%%%%%%%%%%%%%%%%%%%%%%%%%%%%%%%%%%%%%%%%%%%%%%%%%%%%%%%%%%%%%%%%
%
Because of helicity conservation
for (massless) fermions in QED, the dilepton system coupled to a photon
in the process $\mathrm{J}/\psi \rightarrow \gamma^{*} \rightarrow
\ell^+ \ell^-$ has angular momentum projection $\pm 1$ along
$z^\prime$, i.e.\ it can be represented as an eigenstate of
$J_{z^\prime}$, $|\ell^+ \ell^-; 1, l^\prime \rangle$ with $l^\prime =
+ 1$ or $-1$.  We want to express this state as a superposition of
eigenstates of $J_z$, $|\ell^+ \ell^-; 1, l \rangle$ with $l = 0, \pm
1$. To perform this change of quantization axis, we use a general
result of angular momentum theory, which we recall in the following
paragraphs.

We indicate by $R(\alpha, \beta, \gamma)$ the rotation from a generic
set of axes $(x,y,z)$ to the set $(x^\prime,y^\prime,z^\prime)$, $\alpha$,
$\beta$ and $\gamma$ denoting the Euler angles.
%: $R$ can be expressed as a rotation of $\alpha$ around the $z$ axis,
%bringing the new axes to the position $(x_1,y_1,z)$, followed by a
%rotation of $\beta$ about the $y_1$ axis, up to the position
%$(x_1^\prime,y_1,z^\prime)$, and, finally, by a rotation of $\gamma$
%around the $z^\prime$ axis.
Positive rotations are defined by the
right-hand rule. An eigenstate $|J, M^\prime \rangle$ of $J_{z^\prime}$
can then be expressed as a superposition of the eigenstates $|J, M \rangle$ of
$J_z$ through the rotation transformation~\cite{bib:BrinkSatchler}
\begin{equation}
|J, M^\prime \rangle = \sum_{M = -J}^{+J} \mathcal{D}_{M M^\prime}^{J}(R) \,
|J, M \rangle \, . \label{eq:ang_mom_rotation}
\end{equation}
The (complex) rotation matrix elements $\mathcal{D}_{M M^\prime}^{J}$ are
defined as
\begin{equation}
\mathcal{D}_{M M^\prime}^{J}(\alpha, \beta, \gamma) = e^{-i M \alpha } d_{M
M^\prime}^{J}(\beta) e^{-i M^\prime \gamma} \label{eq:D_matrix}
\end{equation}
in terms of the (real) reduced matrix elements
\begin{eqnarray}
&& d_{M M^\prime}^J(\beta)  =
\sum_{t=\max(0,M-M^\prime)}^{\min(J+M,J-M^\prime)}(-1)^{t} \nonumber \\[2mm]
&&\times \frac{\sqrt{(J+M)!\, (J-M)!\, (J+M^\prime)!\, (J-M^\prime)!}}
{(J+M-t)!\, (J-M^\prime-t)!\, t! \, (t-M+M^\prime)!}
~~~\label{eq:reduced_d_matrix} \\[2mm]
&& \times \left(\cos\frac{\beta}{2}\right)^{2J+M-M^\prime-2t}
\left(\sin\frac{\beta}{2}\right)^{2t-M+M^\prime}. \nonumber
\end{eqnarray}

\begin{figure}[htb]
\centering
\resizebox{0.6\linewidth}{!}{%
\includegraphics{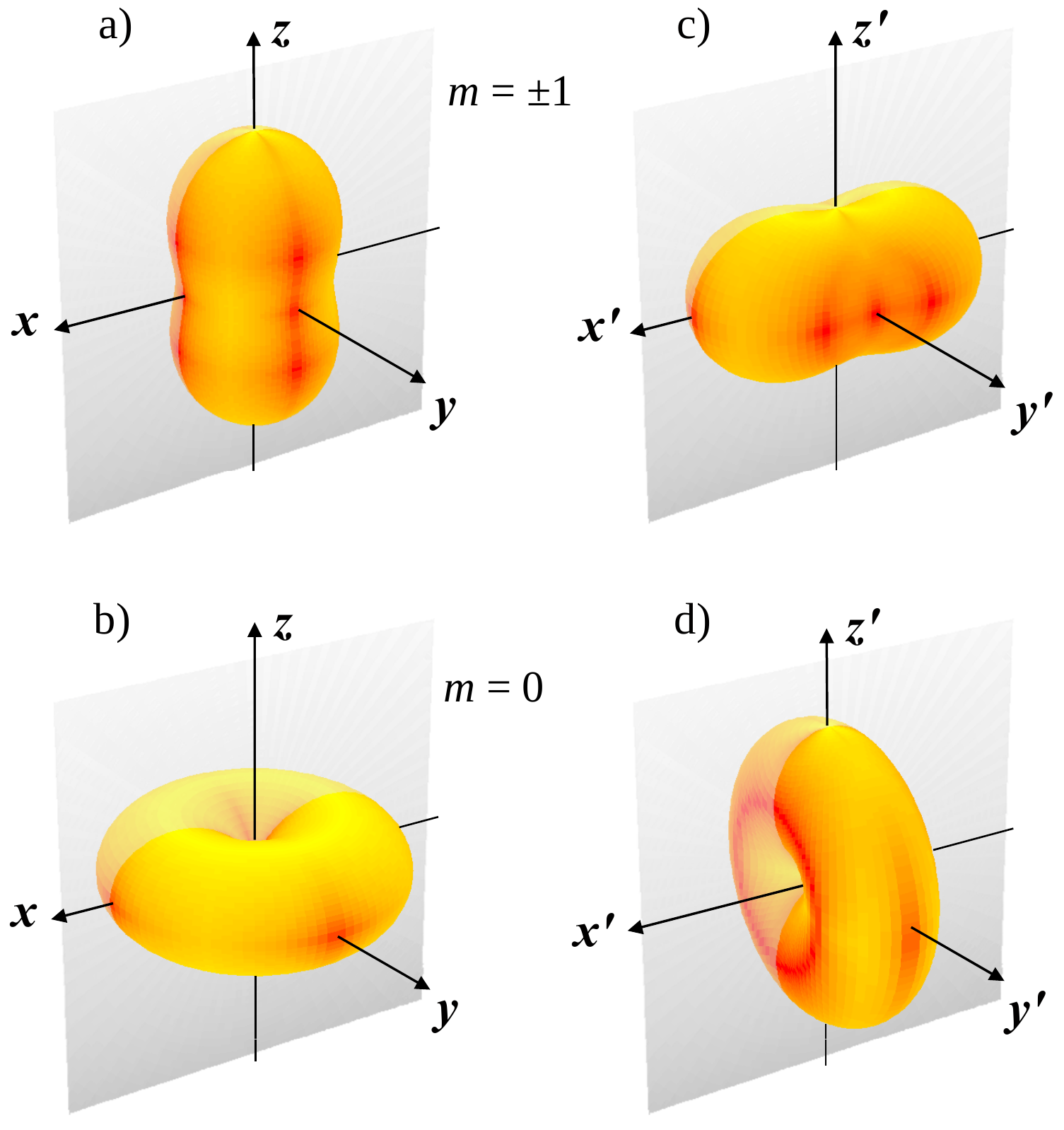}}
\caption{\label{fig:natpols} Graphical representation of the dilepton
  decay distribution of ``transversely'' (a) and ``longitudinally''
  (b) polarized quarkonium in the natural frame and in frames rotated
  by $90^\circ$ (c-d).  The probability of the lepton emission in one
  direction is represented by the distance of the corresponding
  surface point from the origin. }
\end{figure}

The rotation we need in our case has the effect of bringing one
quantization axis ($z$) to coincide with another ($z^\prime$). The
most general rotation performing this projection can be parametrized
with $\beta = \vartheta$ and $\alpha = - \gamma = \varphi$. The
dilepton angular momentum state is therefore expressed in terms of
eigenstates of $J_z$ as
\begin{equation}
|\ell^+ \ell^-; 1, l^\prime \rangle = \sum_{l = 0, \pm 1} \mathcal{D}_{l \,
l^\prime}^{1}(\varphi, \vartheta, - \varphi) \, |\ell^+ \ell^-; 1, l \rangle \,
. \label{eq:dilepton_state}
\end{equation}
The amplitude of the partial process $\mathrm{J}/\psi(m) \rightarrow  \ell^+
\ell^- (l^\prime)$ represented in Fig.~\ref{fig:psidecay} is
\begin{eqnarray}
B_{m l^\prime} \; & = & \;
\sum_{l = 0, \pm 1} \mathcal{D}_{l l^\prime}^{1
*}(\varphi, \vartheta, - \varphi) \, \langle \ell^+ \ell^-; 1, l \; | \,
\mathcal{B} \, | \; \mathrm{J}/\psi; 1, m \rangle \nonumber \\[2mm]
& = & \; B \; \mathcal{D}_{m
l^\prime}^{1 *}(\varphi, \vartheta, - \varphi) \, ,
\label{eq:jpsi_to_ll_amplitude}
\end{eqnarray}
where we imposed that the transition operator $\mathcal{B}$ is of the form
$\langle \ell^+ \ell^-; 1, l \; | \, \mathcal{B} \, | \;
\mathrm{J}/\psi; 1, m \rangle = B \, \delta_{m \, l}$ because of
angular momentum conservation, 
%
%where $\mathcal{B}$ is a transition operator such as $\langle \ell^+
%\ell^-; 1, l \; | \, \mathcal{B} \, | \; \mathrm{J}/\psi; 1, m \rangle
%= B \, \delta_{m \, l}$ (angular momentum conservation),
%
with $B$ independent of $m$ (for rotational
invariance). The total amplitude for $\mathrm{J}/\psi \rightarrow
\ell^+ \ell^- (l^\prime)$, where the \jpsi\ is given by the
superposition written in Eq.~\ref{eq:state}, is
\begin{eqnarray}
B_{l^\prime} \; & = & \; \sum_{m = -1, +1} b_m B \; \mathcal{D}_{m l^\prime}^{1
*}(\varphi, \vartheta, - \varphi) \nonumber \\[2mm]
& = & \; \sum_{m = -1, +1} a_m \;
\mathcal{D}_{m l^\prime}^{1
*}(\varphi, \vartheta, - \varphi) \, . 
\label{eq:jpsi_to_ll_fullamplitude}
\end{eqnarray}
The probability of the transition is obtained by squaring
Eq.~\ref{eq:jpsi_to_ll_fullamplitude} and summing over the
(unobserved) spin alignments ($l^\prime = \pm 1$) of the dilepton
system, with equal weights attributed, for parity conservation, to the
two configurations.  Using Eq.~\ref{eq:D_matrix}, with $d^1_{0, \pm 1}
= \pm \sin \vartheta / \sqrt{2}$, $d^1_{\pm 1, \pm 1} = (1 + \cos
\vartheta) / 2$ and $d^1_{\pm 1, \mp 1} = (1 - \cos \vartheta) / 2$
(from Eq.~\ref{eq:D_matrix}), one obtains the angular distribution
\begin{eqnarray}
  W(\cos \vartheta, \varphi)  && \propto \; \sum_{l^\prime = \pm 1} | B_{l^\prime} |^2 %\nonumber \\
                             \; \propto \; \frac{\mathcal{N}}{(3 +
\lambda_{\vartheta})}\; (1 + \lambda_{\vartheta} \cos^2 \vartheta \nonumber \\[2mm]
&& + \;\lambda_{\varphi} \sin^2 \vartheta \cos 2 \varphi\; +\;
\lambda_{\vartheta \varphi} \sin 2 \vartheta \cos \varphi  
\label{eq:ang_distr_subproc} \\[2mm]
&& + \;\lambda^{\bot}_{\varphi} \sin^2 \vartheta \sin 2 \varphi \;+\;
\lambda^{\bot}_{\vartheta \varphi} \sin 2 \vartheta \sin \varphi  ) \, ,  \nonumber
\end{eqnarray}
with $\mathcal{N} = |a_0|^2 + |a_{+1}|^2 + |a_{-1}|^2$ and
\begin{eqnarray}
  \lambda_{\vartheta} & = & \frac{{\mathcal{N}}-3 |a_0|^2}{\mathcal{N}+|a_0|^2}  \, , \nonumber \\[2mm]
  \lambda_{\varphi}  & = & \frac{ 2 \, \mathrm{Re} [a_{+1}^{(i)*}
    a_{-1}] }{\mathcal{N}+|a_0|^2} \, , \nonumber \\[2mm]
  \lambda_{\vartheta \varphi} &  = & \frac{ \sqrt{2} \, \mathrm{Re} [ a_{0}^{(i)*} ( a_{+1} - a_{-1})] }{\mathcal{N}+|a_0|^2} \,
  , \label{eq:lambdas_vs_amplitudes} \\[2mm]
  \lambda^{\bot}_{\varphi}  & = & \frac{ -2 \, \mathrm{Im} [a_{+1}^{*} a_{-1}] }{\mathcal{N}+|a_0|^2} \, , \nonumber \\[2mm]
  \lambda^{\bot}_{\vartheta \varphi} &  = & \frac{ - \sqrt{2} \,
    \mathrm{Im} [a_{0}^{*} (a_{+1} + a_{-1})] }{\mathcal{N}+|a_0|^2} \, . \nonumber
\end{eqnarray}
Figure~\ref{fig:natpols} shows the shapes of the dilepton decay
distributions in the two polarization cases $m = \pm 1$~(a) and $m =
0$~(b); the $m = +1$ and $m = -1$ configurations are indistinguishable
because of rotation invariance.  The same distributions are also
shown, in panels (c) and (d), as seen when studied in frames rotated
by $90^\circ$, anticipating the discussion in
Section~\ref{sec:frame_dependence}.

%
%%%%%%%%%%%%%%%%%%%%%%%%%%%%%%%%%%%%%%%%%%%%%%%%%%%%%%%%%%%%%%%%%%%%%%%%%%%%%%%%%%%%%%%
%%%%%%%%%%%%%%%%%%%%%%%%%%%%%%%%%%%%%%%%%%%%%%%%%%%%%%%%%%%%%%%%%%%%%%%%%%%%%%%%%%%%%%%
%

It is worth noticing that it is impossible to chose the decay
amplitudes $a_m$ and, therefore, the component amplitudes $b_m$
such that all decay parameters in Eq.~\ref{eq:ang_distr_subproc}
vanish. This means that the angular distribution of the decay of a
$J=1$ state is \emph{never intrinsically isotropic}.  Even if it is
conceivable that a lucky superposition of different production
processes might lead to a fortuitous cancellation of all decay
parameters, such an exceptional case would signal a non-trivial
physical polarization scenario, caused by spin randomization effects,
or (semi-)exclusive configurations in which the observed state is
produced together with certain final state objects. In other words,
polarization is an essential property of the quarkonium states.  This
remark is particularly relevant when we consider that all existing
Monte Carlo generators use an isotropic dilepton distribution as the
default option for quarkonium production in hadronic collisions, a
non-trivial assumption with a strong influence on the acceptance
estimates and, therefore, on both normalizations and kinematic
dependencies of the measured quarkonium cross sections.

In this paper we only consider inclusive production. Therefore, the
only possible experimental definition of the $xz$ plane coincides with
the production plane, containing the directions of the colliding
particles and of the decaying particle itself. The last two terms in
Eq.~\ref{eq:ang_distr_subproc} introduce an asymmetry of the
distribution by reflection with respect to the production plane, an
asymmetry which is not forbidden in individual (parity-conserving)
events. In hadronic collisions, due to the intrinsic parton transverse
momenta, for example, the ``natural'' polarization plane does \emph{not}
coincide event-by-event with  the experimental production plane.
However, the symmetry by reflection must be a property of the observed
\emph{event distribution} when only parity-conserving processes
contribute.
Indeed, the terms in $\sin^2 \vartheta \sin 2 \varphi$ and $\sin 2
\vartheta \sin \varphi$ are unobservable, because they vanish on
average. 

In the presence of $n$ contributing production processes with
weights $f^{(i)}$, the most general \emph{observable} distribution can
be written as
\begin{eqnarray}
  W(\cos \vartheta, \varphi) \, & = & \, \sum_{i = 1}^{n} f^{(i)}
  W^{(i)}(\cos \vartheta, \varphi) \nonumber \\[2mm]
  & \propto & \, \frac{1}{(3 + \lambda_{\vartheta})} \,
  (1 + \lambda_{\vartheta} \cos^2 \vartheta \label{eq:observable_ang_distr} \\[2mm]
  & + &  \lambda_{\varphi} \sin^2 \vartheta \cos 2 \varphi +
  \lambda_{\vartheta \varphi} \sin 2 \vartheta \cos \varphi ) \, ,\nonumber
\end{eqnarray}
where $W^{(i)}(\cos \vartheta, \varphi)$ is the ``elementary'' decay
distribution corresponding to a single subprocess (given by
Eqs.~\ref{eq:ang_distr_subproc} and~\ref{eq:lambdas_vs_amplitudes},
adding the index $(i)$ to the decay
parameters) and each of the three observable shape parameters, $X =
\lambda_{\vartheta}$, $\lambda_{\varphi}$ and $\lambda_{\vartheta
  \varphi}$, is a weighted average of the corresponding parameters,
$X^{(i)}$, characterizing the single subprocesses,
\begin{equation}
  X \, = \, \sum_{i = 1}^{n} \frac{f^{(i)} \mathcal{N}^{(i)}}{3 +
    \lambda_{\vartheta}^{(i)}} \,
  X^{(i)} \left/ \sum_{i = 1}^{n} \frac{f^{(i)} \mathcal{N}^{(i)}}{3 +
      \lambda_{\vartheta}^{(i)}} \right. \, .
\label{eq:parameters}
\end{equation}

We conclude this section with the derivation of formulae which can be used for
the determination of the parameters of the observed angular
distribution, as an alternative to a multi-parameter fit to the
function in Eq.~\ref{eq:observable_ang_distr}. The integration over
either $\varphi$ or $\cos \vartheta$ leads to one-dimensional angular
distributions,
\begin{eqnarray}
  W(\cos\vartheta) \; & \propto & \; \frac{1}{3 + \lambda_{\vartheta}} \left( 1 +
    \lambda_\vartheta \cos^2\vartheta \right) \, ,
\label{eq:costh_distr} \\
W(\varphi) \; & \propto & \; 1 + {\frac{2 \lambda_\varphi}{3
+ \lambda_\vartheta}} \cos 2 \varphi \, , \label{eq:phi_distr}
\end{eqnarray}
from which $\lambda_\vartheta$ and $\lambda_\varphi$ can be determined
in two separate steps, possibly improving the stability of the fit
procedures in low-statistics analyses. The ``diagonal'' term,
$\lambda_{\vartheta \varphi}$, vanishes in both integrations but can
be extracted, for example, by defining the variable $\tilde{\varphi}$
as
\begin{equation}
\tilde{\varphi} = \left\{ \begin{array}{rcl}
\varphi - \frac{3}{4} \pi & \quad \mbox{for} & \cos \vartheta < 0 \\
\varphi - \frac{\pi}{4}   & \quad \mbox{for} & \cos \vartheta > 0 \\
\end{array}
\right.
\end{equation}
(adding or subtracting $2 \pi$ when $\tilde{\varphi}$ does not fall
into one continuous range, e.g.\ $[0, 2 \pi]$) and measuring the
distribution
\begin{equation}
  W(\tilde{\varphi}) \; \propto\; 1 +
    {\frac{\sqrt{2} \, \lambda_{\vartheta\varphi}}{3 +
        \lambda_\vartheta}} \, \cos \tilde{\varphi} \, .
\label{eq:phithdistr}
\end{equation}
Each of the three parameters can also be expressed in terms of an
asymmetry between the populations of two angular topologies (which are
equiprobable only in the unpolarized case):
\begin{eqnarray}
&& \frac{P(|\cos \vartheta| > 1/2) - P(|\cos \vartheta| < 1/2)}{P(|\cos \vartheta|
> 1/2) + P(|\cos \vartheta| < 1/2)} \; = \; \frac{3}{4}
\, \frac{\lambda_\vartheta}{3+\lambda_\vartheta} \, , \nonumber \\[2mm]
&& \frac{P(\cos 2 \varphi > 0) - P(\cos 2 \varphi < 0)}{P(\cos 2 \varphi > 0) +
P(\cos 2 \varphi < 0)} \; = \; \frac{2}{\pi} \, \frac{2
\lambda_\varphi}{3+\lambda_\vartheta} \, , \label{eq:asymmetries} \\[2mm]
&& \frac{P(\sin 2 \vartheta \cos \varphi > 0) - P(\sin 2 \vartheta \cos \varphi <
0)}{P(\sin 2 \vartheta \cos \varphi
> 0) + P(\sin 2 \vartheta \cos \varphi < 0)} \; = \; \frac{2}{\pi}
\, \frac{2
\lambda_{\vartheta \varphi}}{3+\lambda_\vartheta} \, . \nonumber
\end{eqnarray}
%\frac{P(|\cos \vartheta| > 1/2) - P(|\cos \vartheta| < 1/2)}{P(|\cos \vartheta|
%> 1/2) + P(|\cos \vartheta| < 1/2)} \quad & \; = \; \frac{3}{4}
%\, \frac{\lambda_\vartheta}{3+\lambda_\vartheta} \,
%, \\
%\frac{P(\cos 2 \varphi > 0) - P(\cos 2 \varphi < 0)}{P(\cos 2 \varphi > 0) +
%P(\cos 2 \varphi < 0)} \quad \quad & \; = \; \frac{2}{\pi} \, \frac{2
%\lambda_\varphi}{3+\lambda_\vartheta} \,
%, \\
%\frac{P(\sin 2 \vartheta \cos \varphi > 0) - P(\sin 2 \vartheta \cos \varphi <
%0)}{P(\sin 2 \vartheta \cos \varphi
%> 0) + P(\sin 2 \vartheta \cos \varphi < 0)} & \; = \; \frac{2}{\pi}
%\, \frac{2
%\lambda_{\vartheta \varphi}}{3+\lambda_\vartheta} \, . \label{eq:asymmetries}
%\end{split}
%\end{align}
%
In analyses applying efficiency corrections to the reconstructed
angular spectra, the use of these formulae may require an iterative
re-weighting of the Monte Carlo data, in order to compensate for the
effect of the non-uniformity of those experimental corrections. In
``ideal'' experiments with uniform acceptance and efficiencies over
$\cos \vartheta$ and $\varphi$ (such as in Monte Carlo studies at the
generation level) the parameters can be obtained from the average values of
certain angular distributions:
\begin{align}
\begin{split}
\langle \cos^2 \vartheta \rangle  \quad & \; = \; \frac{1 + \frac{3}{5}
\lambda_\vartheta}{3 + \lambda_\vartheta} \,
, \\
\langle \cos 2 \varphi \rangle  \quad & \; = \; \frac{\lambda_\varphi}{3 +
\lambda_\vartheta} \,
, \\
\langle \sin 2 \vartheta \cos \varphi \rangle  & \; = \; \frac{4}{5} \,
\frac{\lambda_{\vartheta \varphi}}{3 + \lambda_\vartheta} \, .
\label{eq:means}
\end{split}
\end{align}

%%%%%%%%%%%%%%%%%%%%%%%%%%%%%%%%%%%%%%%%%%%%%%%%%%%%%%%%%%%%%%%%%%%%%%%%%
\section{Dependence of the measurement on the observation frame}
\label{sec:frame_dependence}

\begin{figure}[ht]
\centering
\resizebox{0.9\linewidth}{!}{%
\includegraphics{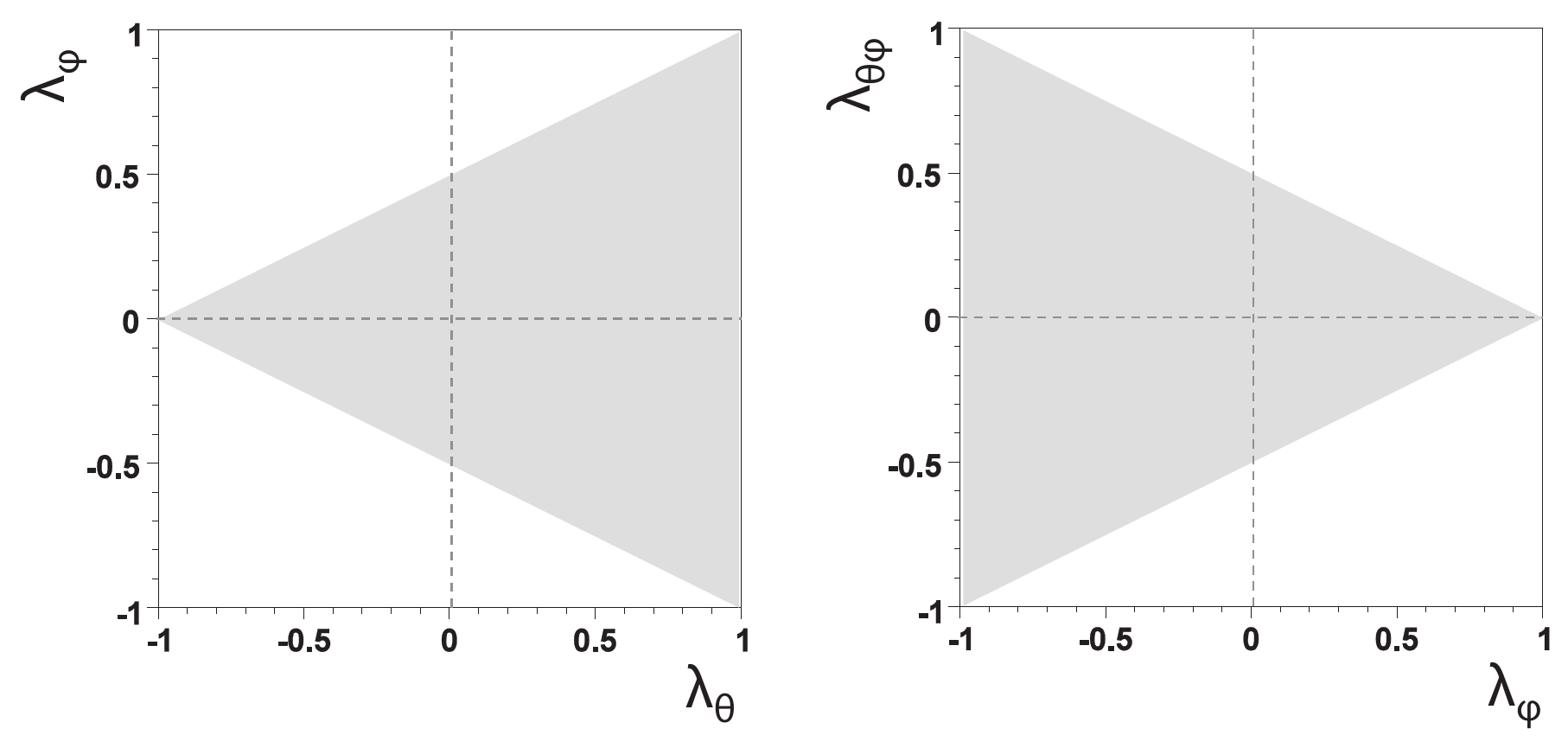}}
\caption{\label{fig:triangles} Allowed regions for the decay angular
parameters. }
\end{figure}

All possible \emph{experimentally definable} polarization axes in inclusive
measurements belong to the production plane (defined in Fig.~\ref{fig:frames}). We can,
therefore, parametrize the transformation from an observation frame to another
by one angle describing a rotation about the $y$ axis. Instead of rotating
the angular momentum state vectors, we can apply a purely geometrical
transformation directly to the observable angular distribution. 
The rotation matrix
\begin{equation}
R_y(\delta) = \begin{pmatrix} \cos \delta & 0 & -\sin \delta \\
                                     0    & 1 &     0       \\
                             \sin \delta & 0 & \cos \delta \\
              \end{pmatrix}
\end{equation}
brings the old frame to coincide with the new one, the positive
sign of $\delta$ being defined by the right-hand rule (we will discuss
in Section~\ref{sec:kinematics} how the sign of $\delta$ depends in an
observable way on the conventions chosen for the orientation of the $z$ and $y$
axes, and how, specifically, the angle between the HX and CS axes depends on the
quarkonium production kinematics). The unit vector $\hat{r} = (\sin \vartheta
\cos \varphi, \sin \vartheta \sin \varphi, \cos \vartheta)$ indicating the
lepton direction in the old frame is then expressed as $\hat{r} =
R^{-1}_y(\delta) \hat{r^\prime}$ as a function of the coordinates in the new
frame. In particular,
\begin{equation}
\cos \vartheta \; = \; - \sin \delta \sin \vartheta^\prime \cos \varphi^\prime
\; + \; \cos \delta \cos \vartheta^\prime \, .
\label{eq:costh_transform}
\end{equation}
Substituting Eq.~\ref{eq:costh_transform} into
Eq.~\ref{eq:observable_ang_distr}, we obtain the angular distribution in the
rotated frame:
\begin{eqnarray}
  W^\prime(\cos \vartheta^\prime, \varphi^\prime)  \; & \propto & \;
  \frac{1}{3 + \lambda_{\vartheta}^\prime} \,
  (1 \, + \lambda^\prime_{\vartheta} \cos^2 \vartheta^\prime 
  \label{eq:ang_distr_rotated} \\[2mm]
  & + & \lambda^\prime_{\varphi} \sin^2 \vartheta^\prime \cos 2 \varphi^\prime
  \, + \lambda^\prime_{\vartheta \varphi} \sin 2 \vartheta^\prime \cos
  \varphi^\prime)  \, , \nonumber
\end{eqnarray}
where
\begin{equation}
\begin{split}
\lambda^{\prime}_\vartheta & =  \frac{{\lambda_\vartheta
 - 3\Lambda }}{{1 + \Lambda }}\, ,  \quad
\lambda^{\prime}_\varphi  =  \frac{{\lambda_\varphi
 + \Lambda }}{{1 + \Lambda }}\, ,  \\[2mm]
\lambda^{\prime}_{\vartheta \varphi} & =  \frac{{\lambda_{\vartheta \varphi}
\cos 2\delta  - \frac{1}{2}\, (\lambda_\vartheta - \lambda_\varphi )
\sin 2\delta }} {{1 + \Lambda }}\, , \\[2mm]
\mathrm{with} \quad \Lambda & = \frac{1}{2}\, (\lambda_\vartheta -
\lambda_\varphi)\sin^2 \delta - \frac{1}{2}\, \lambda_{\vartheta \varphi} \sin
2\delta  \, . \label{eq:lambda_transf}
\end{split}
\end{equation}
Since the magnitude of the ``polar anisotropy'', $\lambda_\vartheta$, never
exceeds $1$ in any frame, we deduce the frame-independent inequalities
\begin{equation}
|\lambda_\varphi| \le \frac{1}{2}\, (1 + \lambda_\vartheta ) \, , \quad
|\lambda_{\vartheta \varphi}| \le \frac{1}{2}\, (1 - \lambda_\varphi ) \, ,
\label{eq:triangles}
\end{equation}
which imply the bounds $|\lambda_\varphi| \le 1 $ and $|\lambda_{\vartheta
\varphi}| \le 1$. More interestingly, we can see that $|\lambda_\varphi| \le
0.5$ when $\lambda_{\vartheta} = 0$ and must vanish when $\lambda_{\vartheta}
\to -1$. The most general phase space for the three angular parameters is
represented in Fig.~\ref{fig:triangles}.
There is an alternative notation, widespread in the literature, where
the coefficients $\lambda$, $\nu/2$ and $\mu$ replace, respectively,
$\lambda_{\vartheta}$, $\lambda_{\varphi}$ and $\lambda_{\vartheta
  \varphi}$.  In that case, hence, we have $|\nu| \le 2$.

%
%%%%%%%%%%%%%%%%%%%%%%%%%%%%%%%%%%%%%%%%%%%%%%%%%%%%%%%%%%%%%%%%%%%%%%%%%%%%%%%%%%%%%%%
%%%%%%%%%%%%%%%%%%%%%%%%%%%%%%%%%%%%%%%%%%%%%%%%%%%%%%%%%%%%%%%%%%%%%%%%%%%%%%%%%%%%%%%
%

To illustrate the importance of the choice of the observation frame,
we consider specific examples assuming, for simplicity, that the
observation axis is perpendicular to the natural axis ($\delta = \pm
90^\circ$). This case is of physical relevance since when the
decaying particle is produced with small longitudinal momentum
($|p_{\rm L}| \ll p_{\rm T}$, a frequent kinematic configuration in
collider experiments) the CS and HX frames are actually perpendicular
to one another. When $\delta = 90^\circ$, a natural ``transverse''
polarization ($\lambda_\vartheta = +1$ and 
$\lambda_\varphi = \lambda_{\vartheta\varphi} = 0$), for example, transforms
(Eq.~\ref{eq:lambda_transf}) into an observed polarization of opposite
sign (but not fully ``longitudinal''), $\lambda^\prime_\vartheta =
-1/3$, with a significant azimuthal anisotropy,
$\lambda^\prime_\varphi = 1/3$, shown in Fig.~\ref{fig:natpols}~(c). 
In terms of
angular momentum wave functions, a state which is fully ``transverse''
with respect to one quantization axis is a coherent superposition of
50\% ``transverse'' and 50\% ``longitudinal'' components with respect
to an axis rotated by $90^\circ$ (Eq.~\ref{eq:ang_mom_rotation}):
\begin{equation}
|1, \pm 1 \rangle \quad \xrightarrow{ 90^\circ } \quad \frac{1}{2} \;
|1, +1 \rangle \; + \; \frac{1}{2} \; |1, -1 \rangle \; \mp \;
\frac{1}{\sqrt{2}} \; |1, 0 \rangle  \, .
\end{equation}
The amplitude of the transition of this mixed state to the ``rotated'' dilepton
state in Eq.~\ref{eq:dilepton_state} contains three terms with \emph{relative}
phases (due to the $\varphi$ dependence of the rotation matrix) giving rise to
the observable azimuthal dependence. The same polar anisotropy
$\lambda^\prime_\vartheta = -1/3$ would be measured in the presence of a
mixture of \emph{at least two different processes} resulting in 50\%
``transverse'' and 50\% ``longitudinal'' natural polarization along the chosen
axis. In this case, however, no azimuthal anisotropy would be
observed.
As a second example, we note that a fully ``longitudinal'' natural
polarization ($\lambda_\vartheta = -1$) translates, in a frame rotated
by $90^\circ$ with respect to the natural one,
Fig.~\ref{fig:natpols}~(d), into a fully ``transverse'' polarization
($\lambda^\prime_\vartheta = +1$), accompanied by a maximal azimuthal
anisotropy ($\lambda^\prime_\varphi = -1$). In terms of angular
momentum, the measurement in the rotated frame is performed on a
coherent admixture of states,
\begin{equation}
|1, 0 \rangle \quad \xrightarrow{\; 90^\circ \;} \quad \frac{1}{\sqrt{2}}
\; |1, +1 \rangle  - \frac{1}{\sqrt{2}} \; |1, -1 \rangle  \, ,
\end{equation}
while a \emph{natural} ``transverse'' polarization would originate
from the statistical superposition of \emph{uncorrelated} $|1, +1
\rangle$ and $|1, -1 \rangle$ states. The two physically very
different cases of a natural transverse polarization observed in the
natural frame, shown in Fig.~\ref{fig:natpols}~(a), and a natural longitudinal
polarization observed in a rotated frame, shown in
Fig.~\ref{fig:natpols}~(d), are
experimentally indistinguishable when the azimuthal anisotropy
parameter is integrated out.  These examples show that a measurement
(or theoretical calculation) consisting only in the determination of
the polar parameter $\lambda_\vartheta$ in one frame contains an
ambiguity which prevents fundamental (model-independent)
interpretations of the results. The polarization is only fully
determined when \emph{both} the polar and the azimuthal components of the
decay distribution are known, or when the distribution is analyzed in
at least two geometrically complementary frames.

%%%%%%%%%%%%%%%%%%%%%%%%%%%%%%%%%%%%%%%%%%%%%%%%%%%%%%%%%%%%%%%%%%%%%%%%%%%%%%%%%%%
\section{Effect of production kinematics on the observed decay kinematics }
\label{sec:kinematics}

Ideally, the dependence of the polarization on the momentum components
of the produced quarkonium should reflect the relative contribution of
individual production processes in different kinematic regimes,
thereby providing information of \emph{fundamental} physical interest.
However, the observations are, in general, affected by some
experimental limitations, which must be carefully taken in
consideration.
First, the frame-dependent polarization parameters
$\lambda_\vartheta$, $\lambda_\varphi$ and $\lambda_{\vartheta
  \varphi}$ can be affected by a strong \emph{explicit} kinematic
dependence (encoded in the parameter $\delta$ in
Eq.~\ref{eq:lambda_transf}), reflecting the change in direction of the
chosen experimental axis (with respect to the ``natural axis'') as a
function of the quarkonium momentum.
Second, detector acceptances and event samples with limited statistics
induce a dependence of the measurement on the distribution of
events effectively accepted by the experimental apparatus.

To better explain the first problem, let us consider the HX and CS
frames as the experimental and natural frames, respectively.  We start
by calculating the angle between the polarization axes of the CS and
HX frames as a function of the quarkonium momentum. The beam momenta
in the ``laboratory'' frame (centre of mass of the colliding
particles), written in longitudinal and transverse components with
respect to the quarkonium direction, are $\vec{P}_1 = -\vec{P}_2 = P
\cos \Theta \; \hat{\imath}_\| + P \sin \Theta \; \hat{\imath}_\bot$,
where $P$ is their modulus and $\Theta$ is the angle formed by the
quarkonium momentum with respect to the beam axis, defined in terms of
the quarkonium momentum $\vec{p}$ as $\cos \Theta = p_\mathrm{L} / p$.
When boosted to the quarkonium rest frame, the two vectors become
(neglecting the masses of the colliding particles) $\vec{P^\prime}_1 =
(\gamma P \cos \Theta - \beta \gamma P) \; \hat{\imath}_\| + P \sin
\Theta \; \hat{\imath}_\bot$ and $\vec{P^\prime}_2 = (- \gamma P \cos
\Theta - \beta \gamma P) \; \hat{\imath}_\| -P \sin \Theta \;
\hat{\imath}_\bot$, where $\gamma = E/m$ is the Lorentz factor
of the quarkonium state, and $\beta = p/E =
\sqrt{1 - 1/\gamma^2}$. The unit vectors indicating the $z$ axis
directions in the HX and CS frames are
\begin{align}
\begin{split}
\hat{z}_{\mathrm{HX}} \; & = \; -\frac{\vec{P^\prime}_1 +
\vec{P^\prime}_2}{|\vec{P^\prime}_1 + \vec{P^\prime}_2|} \; = \;
\frac{\vec{p}}{p} \, , \\[2mm]
\hat{z}_{\mathrm{CS}} \; & = \; \frac{P^\prime_2
\vec{P^\prime}_1 - P^\prime_1 \vec{P^\prime}_2 }{|P^\prime_2 \vec{P^\prime}_1 -
P^\prime_1 \vec{P^\prime}_2|}
  \, . \label{eq:z_axis}
\end{split}
\end{align}
By definition, $\hat{z}_{\mathrm{HX}} = \hat{\imath}_\|$, while
$\hat{z}_{\mathrm{CS}}$ can now be expressed as $\cos \tau \;
\hat{\imath}_\| + \sin \tau \; \hat{\imath}_\bot $, $\tau$ being the
angle between the two axes:
\begin{align}
\begin{split}
\cos \tau \; & = \; \frac{\frac{1}{\gamma} \cos \Theta}{\sqrt{
\frac{1}{\gamma^2} \cos^2 \Theta + \sin^2 \Theta}} \; = \; \frac{m \,
p_\mathrm{L}}{m_\mathrm{T} \, p}
\, , \\
\sin \tau \; & = \; \frac{ \sin \Theta}{\sqrt{ \frac{1}{\gamma^2} \cos^2 \Theta
+ \sin^2 \Theta}} \; = \; \frac{E \, p_\mathrm{T}}{m_\mathrm{T} \, p} \, .
\label{eq:angle_CS_HX}
\end{split}
\end{align}
We see that the result depends only on the momentum and mass of the
quarkonium state ($m_\mathrm{T} = \sqrt{m^2 + p_\mathrm{T}^2}$).  The
angle $\delta$ entering in Eq.~\ref{eq:lambda_transf}, equal to $\tau$
in magnitude, defines the \emph{positive} rotation (respecting the
right-hand rule) from one frame to the other. Its sign depends,
therefore, on the exact conventions used for the orientation of the
axes $y$ and $z$ of the polarization frames. In the convention where
the $y$ axis is defined as
\begin{equation}
\hat{y} \; = \; \frac{( \vec{P^\prime}_1 \times \vec{P^\prime}_2
)}{|\vec{P^\prime}_1 \times \vec{P^\prime}_2|} \label{eq:y_axis}
\end{equation}
and the $z$ axis is defined by Eq.~\ref{eq:z_axis}, with the ``first'' beam
oriented as the laboratory $z$ axis, the positive rotation is the one bringing
the HX axis to coincide with the CS axis. We thus write, using
Eq.~\ref{eq:angle_CS_HX},
\begin{equation}
\delta_{\mathrm{HX} \rightarrow \mathrm{CS}} \; = \; - \delta_{\mathrm{CS}
\rightarrow \mathrm{HX}} \; = \; \arccos \left( 
\frac{m\, p_\mathrm{L}}{m_\mathrm{T}\, p}
 \right) \, . \label{eq:delta_HX_to_CS}
\end{equation}
Equation~\ref{eq:lambda_transf}, containing terms of the form
\begin{align}
\begin{split}
\sin^2\delta_{\mathrm{HX} \rightarrow \mathrm{CS}} \; = \;
\sin^2\delta_{\mathrm{CS} \rightarrow \mathrm{HX}} \; & = \quad
\frac{p_\mathrm{T}^2 \, E^2}{p^2 \, m_\mathrm{T}^2} \, , \\
\sin 2\delta_{\mathrm{HX} \rightarrow \mathrm{CS}} \; = \; - \sin
2\delta_{\mathrm{CS} \rightarrow \mathrm{HX}} \; & = \; \frac{2 \, m
  \, p_\mathrm{T} \, p_\mathrm{L} \, E}{p^2 \, m_\mathrm{T}^2} \, ,
\label{eq:delta2_HX_to_CS}
\end{split}
\end{align}
is now explicitly seen as a kinematic-dependent transformation.

%
%%%%%%%%%%%%%%%%%%%%%%%%%%%%%%%%%%%%%%%%%%%%%%%%%%%%%%%%%%%%%%%%%%%%%%%%%%%%%%%%%%%%%%%
\begin{figure*}[t!]
\centering
\resizebox{0.73\linewidth}{!}{%
\includegraphics{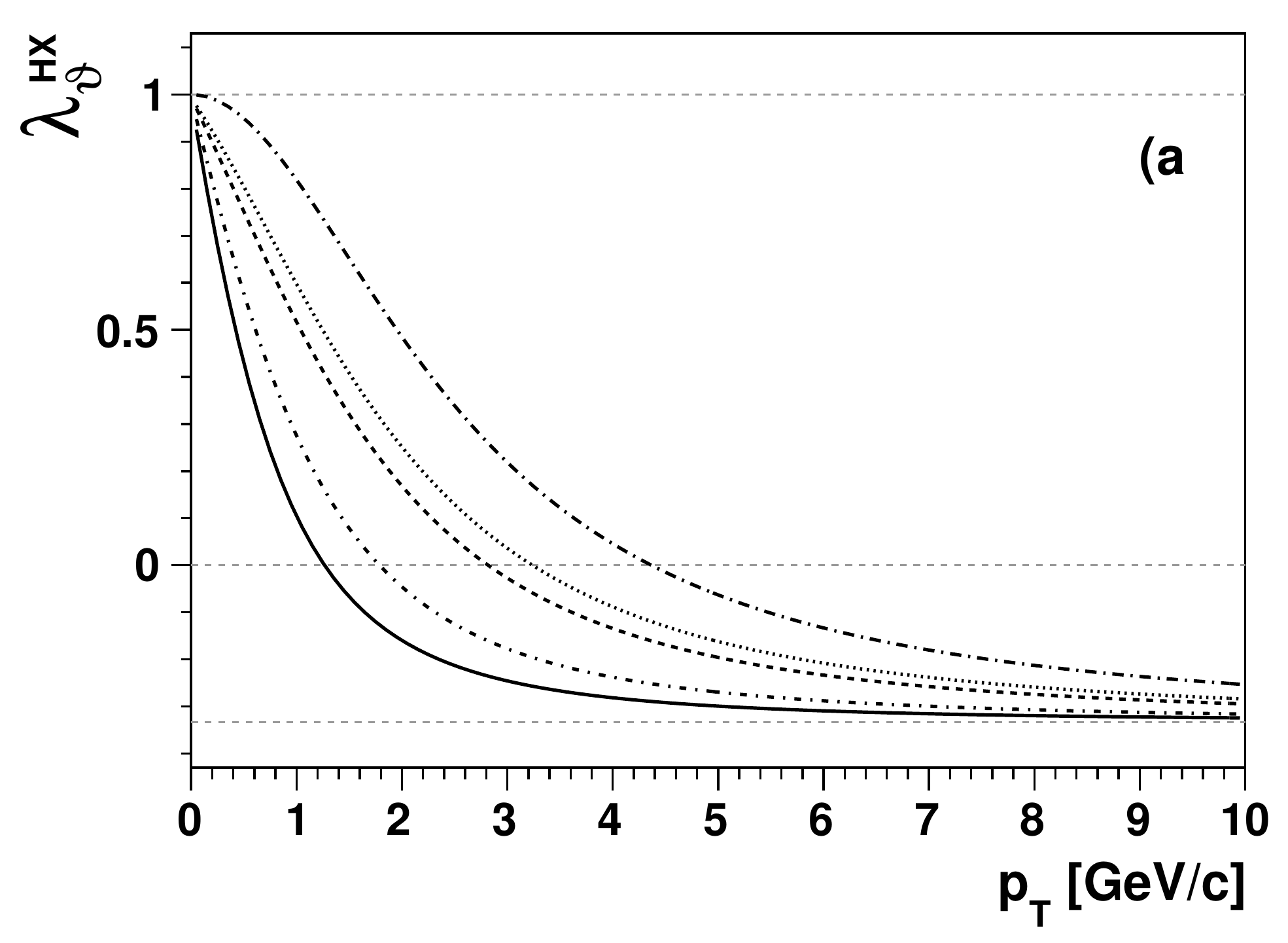}
\includegraphics{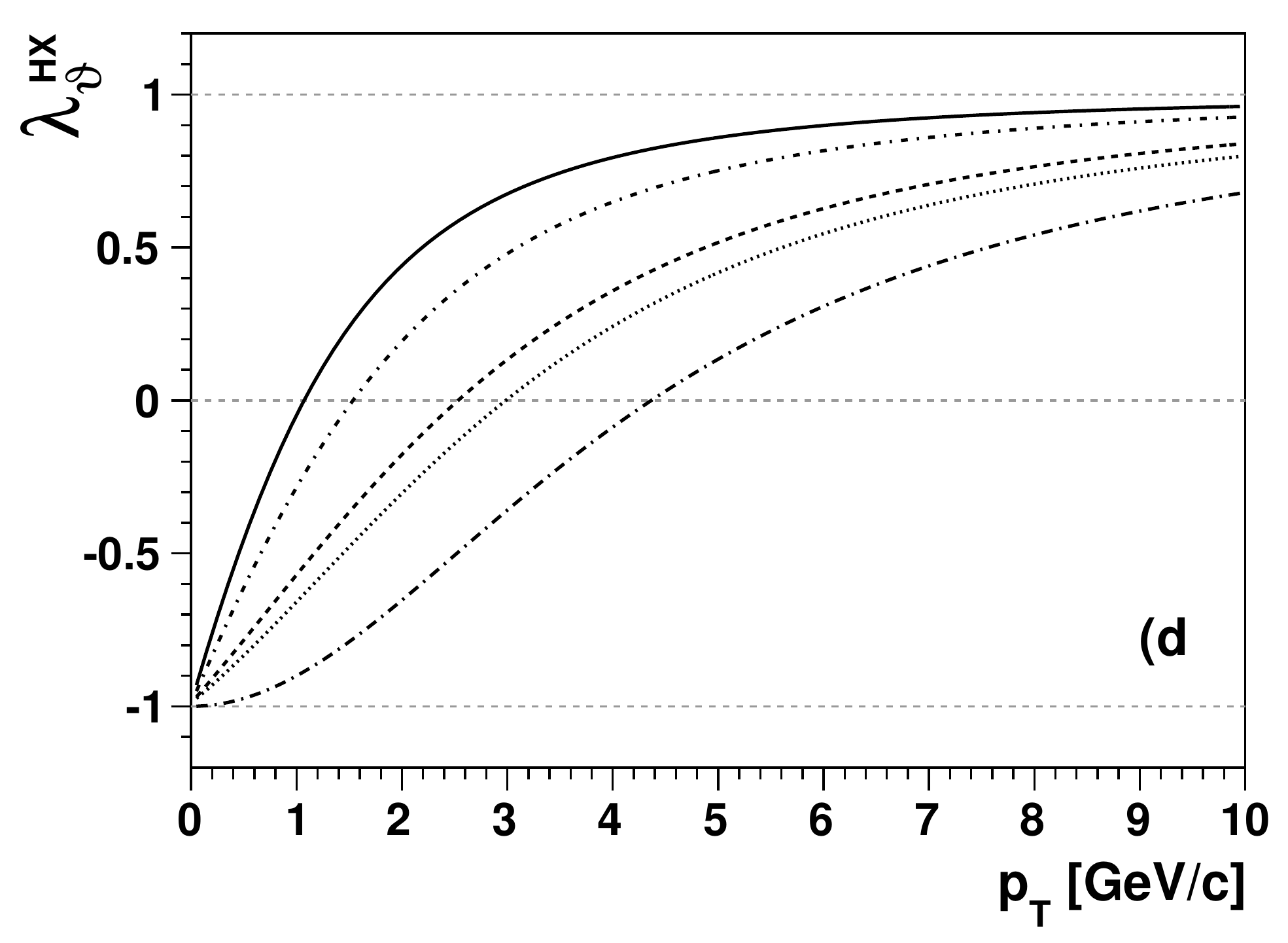}}
\resizebox{0.73\linewidth}{!}{%
\includegraphics{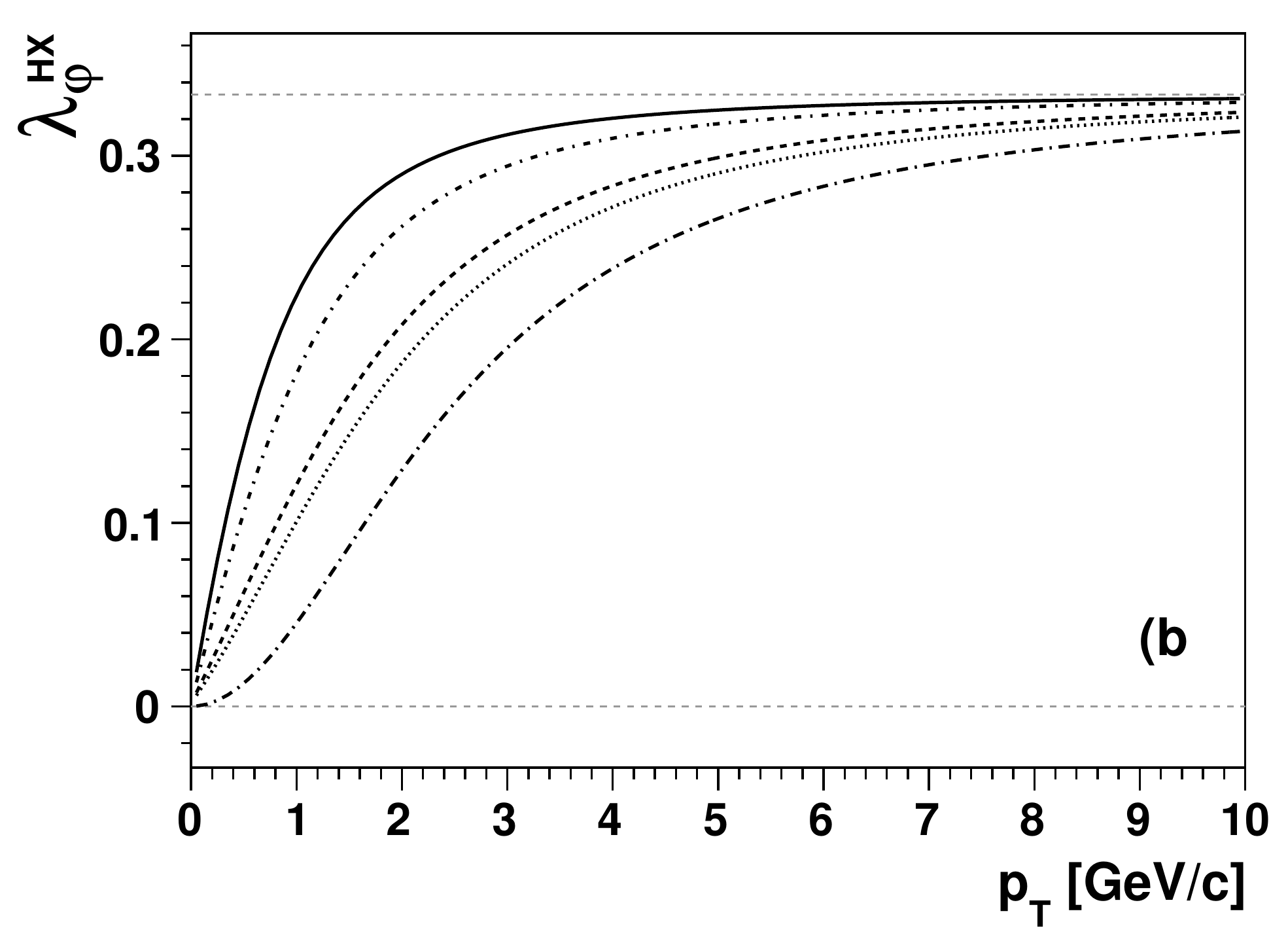}
\includegraphics{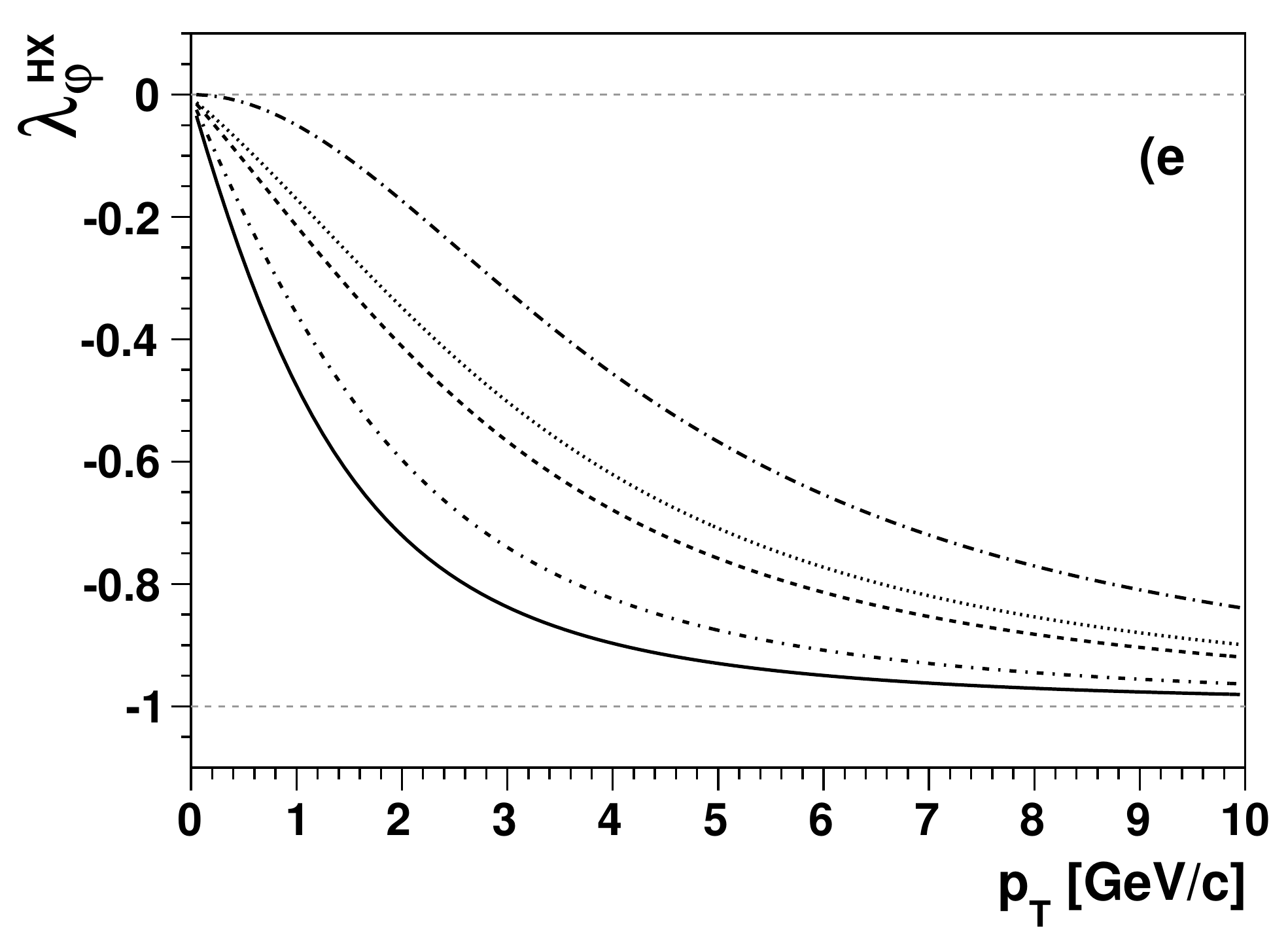}}
\resizebox{0.73\linewidth}{!}{%
\includegraphics{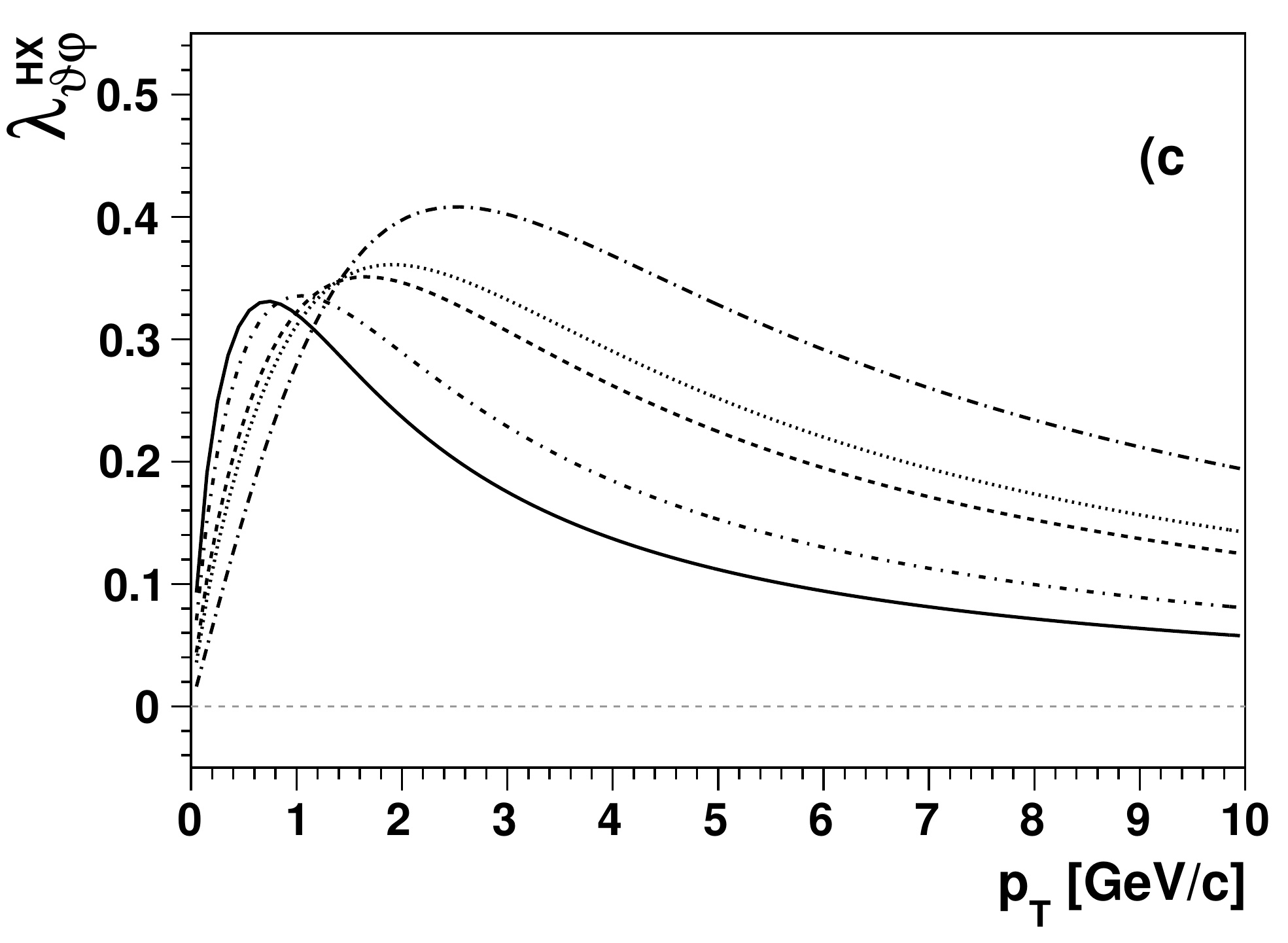}
\includegraphics{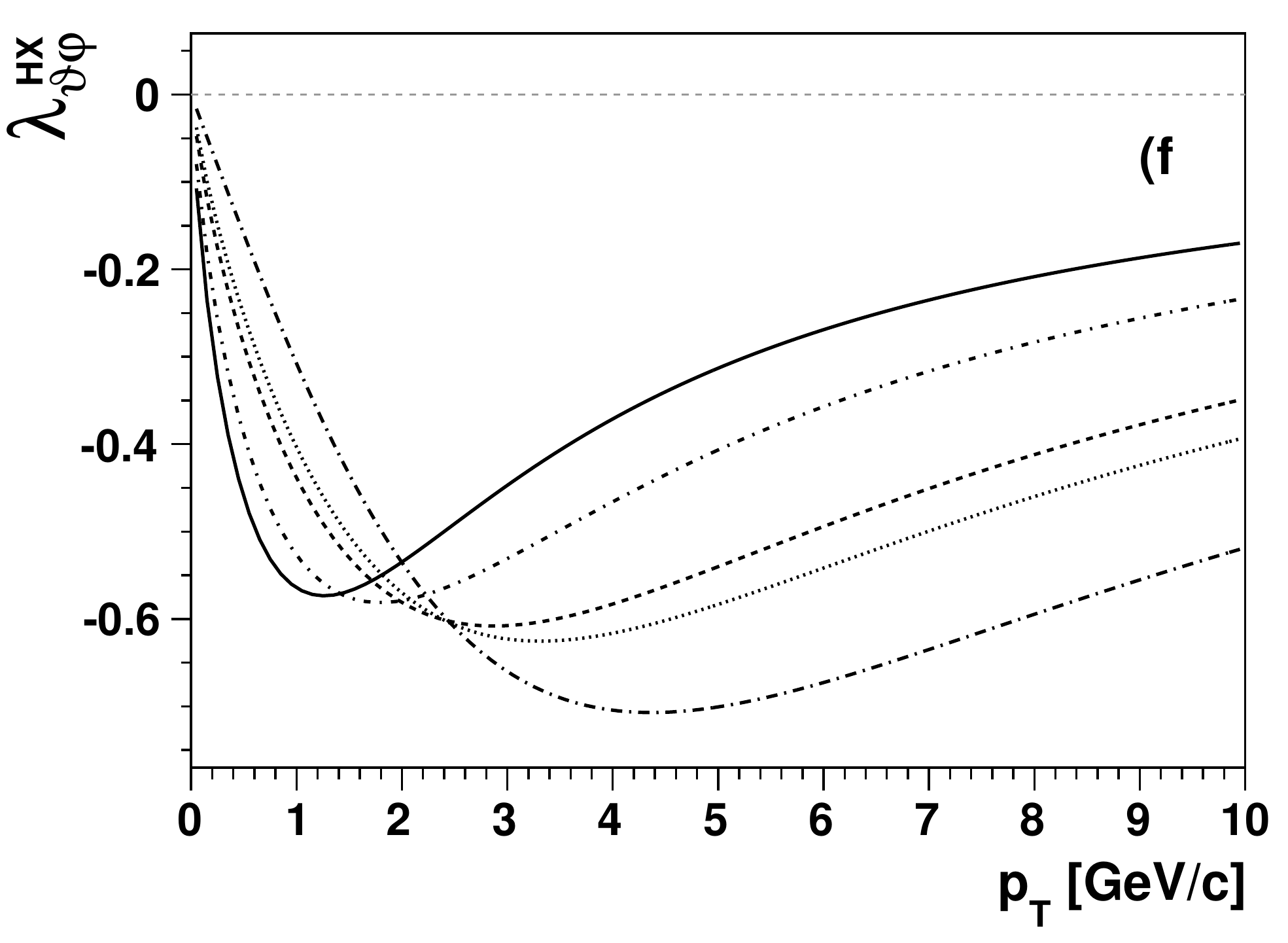}}
\caption{Kinematic dependence of the \jpsi\
%\upsOneS\
  decay angular
  distribution seen in the HX frame, for natural polarizations
  $\lambda_{\vartheta}$\,$=$\,$+1$ (a-b-c) and
  $\lambda_{\vartheta}$\,$=$\,$-1$ (d-e-f) in the CS frame. The curves
  correspond to different rapidity intervals; from the solid line:
  $|y| < 0.6$ (CDF), $|y| < 0.9$ (ALICE), $|y| < 1.8$ (D0), $|y| <
  2.5$ (ATLAS and CMS), $2< |y| < 5$ (LHCb). For simplicity the event
  populations were generated flat in rapidity. The sign of
  $\lambda_{\vartheta \varphi}$ depends on the definition of the $y$
  axis of the polarization frame, here taken as
  $\mathrm{sign}(p_\mathrm{L}) ( \vec{P^\prime}_1 \times
  \vec{P^\prime}_2 ) /|\vec{P^\prime}_1 \times \vec{P^\prime}_2|$,
  where $\vec{P^\prime}_{1,2}$ are the momenta of the colliding
  particles in the meson's rest frame.}
\label{fig:kindep_lambda}
\end{figure*}
%\includegraphics[width=0.25\linewidth, angle=90]{fig_kindep_lambda_th_11}
%\includegraphics[width=0.25\linewidth, angle=90]{fig_kindep_lambda_th_9}
%\includegraphics[width=0.25\linewidth, angle=90]{fig_kindep_lambda_ph_11}
%\includegraphics[width=0.25\linewidth, angle=90]{fig_kindep_lambda_ph_9}
%\includegraphics[width=0.25\linewidth, angle=90]{fig_kindep_lambda_thph_11}
%\includegraphics[width=0.25\linewidth, angle=90]{fig_kindep_lambda_thph_9}
%\caption{\label{fig:kindep_lambda} The kinematic dependence of the \upsOneS\
%dilepton decay angular distribution observed in the HX frame, when the natural
%polarization is $\lambda_{\vartheta} = +1$ (a-b-c) or $\lambda_{\vartheta} =
%-1$ (d-e-f) in the CS frame. The curves in each plot correspond to the rapidity
%values $y = 0.05$ (continuous line), $0.1$, $0.2$, $0.4$, $0.6$, $1.0$, $2.4$.
%}
%\end{figure}
%%%%%%%%%%%%%%%%%%%%%%%%%%%%%%%%%%%%%%%%%%%%%%%%%%%%%%%%%%%%%%%%%%%%%%%%%%%%%%%%%%%%%%%
%

As an example, we show in Fig.~\ref{fig:kindep_lambda} how natural
%%%\upsAll\ 
\jpsi\ polarizations $\lambda_\vartheta = +1$ and $-1$ in the CS frame
(with $\lambda_{\varphi} = \lambda_{\vartheta \varphi} = 0$ and no
intrinsic kinematic dependence) translate into different \pt-dependent
polarizations measured in the HX frame in different rapidity
acceptance windows, representative of the acceptance ranges of several
Tevatron and LHC experiments.  Corresponding figures for the \upsOneS\
case can be seen in Ref.~\cite{bib:ImprovedQQbarPol}.
The same results, except for a change in the sign of
$\lambda_{\vartheta \varphi}$, the only parameter depending on the
sign of the rotation angle, are obtained if the roles of the two
frames are inter-exchanged.

The change of sign of the rapidity does not change the
$\lambda_{\vartheta}$ and $\lambda_{\varphi}$ curves.  However, the
sign of $\lambda_{\vartheta \varphi}$ can change from
positive to negative rapidity, depending on the convention used for
the orientation of the axes. If the axes are defined as in
Eqs.~\ref{eq:z_axis} and~\ref{eq:y_axis} at both positive and negative
rapidity, always taking as ``first'' beam the one positively oriented
in the laboratory, $\lambda_{\vartheta \varphi}$ (proportional to
$\sin 2 \delta$ with $0 < \delta < \pi$) is forced to change sign when
the rapidity changes sign. Any measurement integrating events over a
range in rapidity where the acceptance is symmetrical around zero
would, therefore, yield $\lambda_{\vartheta \varphi} = 0$. In order to
avoid this cancellation, the axis definitions in Eqs.~\ref{eq:z_axis}
and~\ref{eq:y_axis} can be improved by inverting the orientations of
$\hat{y}$ and $\hat{z}_{\mathrm{CS}}$ for negative rapidity
(correspondingly restricting the domain of the rotation angle to $0 <
\delta < \pi / 2$).

Having seen how the strong kinematic dependence induced by the choice
of the observation frame can mimic and/or mask the fundamental
(``intrinsic'') dependencies reflecting the production mechanisms, let
us now discuss the additional problems caused by common experimental
limitations.
Experiments can only measure the net polarization of the specific
cocktail of quarkonium events accepted by the detector, trigger and
analysis cuts.  Moreover, each measured value necessarily implies an
integration over certain ranges (bins or cells) of the quarkonium
momentum components.  If the polarization depends on the kinematics,
this binning effectively biases the measured angular parameters, in
different ways for experiments having different differential
acceptances.  In other words, two experiments covering the same
kinematic interval can measure different average polarizations.  This
problem can be solved by presenting the results in narrow intervals of
the probed phase space.
Similarly, theoretical calculations aimed at comparisons with
experimental data should consider how the momentum distributions are
distorted by the acceptances of those experiments.  Alternatively, the
predictions should avoid kinematic integrations or, even better, be
provided as event-level information to be embedded in the Monte Carlo
simulations of the experiments.
These considerations provide a further motivation for reporting
measurements and theoretical calculations in frame-independent terms,
as we will discuss in the next section.

%%%%%%%%%%%%%%%%%%%%%%%%%%%%%%%%%%%%%%%%%%%%%%%%%%%%%%%%%%%%%%%%%%%%%%%%%%%%%%%%%%%
\section{A frame-invariant approach}
\label{sec:invariant}

The general frame-transformation relations in
Eq.~\ref{eq:lambda_transf} imply the existence of an invariant
quantity, definable in terms of $\lambda_{\vartheta}$,
$\lambda_{\varphi}$ and $\lambda_{\vartheta \varphi}$, in one of the
following equivalent forms:
\begin{equation}
  \mathcal{F}_{\{c_i\}} \, = \,  \frac{(3 + \lambda_\vartheta) +
    c_1 (1 - \lambda_\varphi)}{c_2 (3 + \lambda_\vartheta) +
    c_3 (1 - \lambda_\varphi)} \, .
\label{eq:invariants}
\end{equation}
An account of the fundamental meaning of the frame-invariance of these
quantities can be found in Ref.~\cite{bib:LTGen}. We will consider
here, specifically, the form
\begin{equation}
  \tilde{\lambda} \, \equiv \,  \mathcal{F}_{\{-3,0,1\}} \,
  = \, \frac{\lambda_\vartheta + 3 \lambda_\varphi }{1 -
    \lambda_\varphi} \, . \label{eq:lambda_tilde}
\end{equation}
In the special case when the observed distribution is the
superposition of $n$ ``elementary'' distributions of the kind $1 +
\lambda_\vartheta^{(i)} \cos^2\vartheta$, with event weights
$f^{(i)}$, with respect to $n$ different polarization axes,
$\tilde{\lambda}$ represents a weighted average of the $n$
polarizations, insensitive to
%%% made \emph{irrespectively} of
the orientations of the corresponding axes:
\begin{equation}
 \tilde{\lambda} \; = \;
\sum_{i = 1}^{n} \frac{f^{(i)}}{3 + \lambda_{\vartheta}^{(i)}}
\, \lambda^{(i)}_{\vartheta}
\, \bigg/ \, \sum_{i = 1}^{n} \frac{f^{(i)}}{3 +
  \lambda_{\vartheta}^{(i)}} \, .
% \frac{ \sum_{i = 1}^{n} \frac{f^{(i)}}{3 +
%     \lambda_{\vartheta}^{(i)}} \lambda^{(i)}_{\vartheta} }{\sum_{i =
%     1}^{n} \frac{f^{(i)}} {3 + \lambda_{\vartheta}^{(i)}}} \, .
\label{eq:lambda_tilde_meaning}
\end{equation}
The determination of an invariant quantity is immune to ``extrinsic''
kinematic dependencies induced by the observation perspective and is,
therefore, less acceptance-dependent than the standard anisotropy
parameters $\lambda_{\vartheta}$, $\lambda_{\varphi}$ and
$\lambda_{\vartheta \varphi}$.

%%%%%%%%%%%%%%%%%%%%%%%%%%%%%%%%%%%%%%%%%%%%
\begin{figure*}[ht]
\centering
\resizebox{0.73\linewidth}{!}{%
\includegraphics{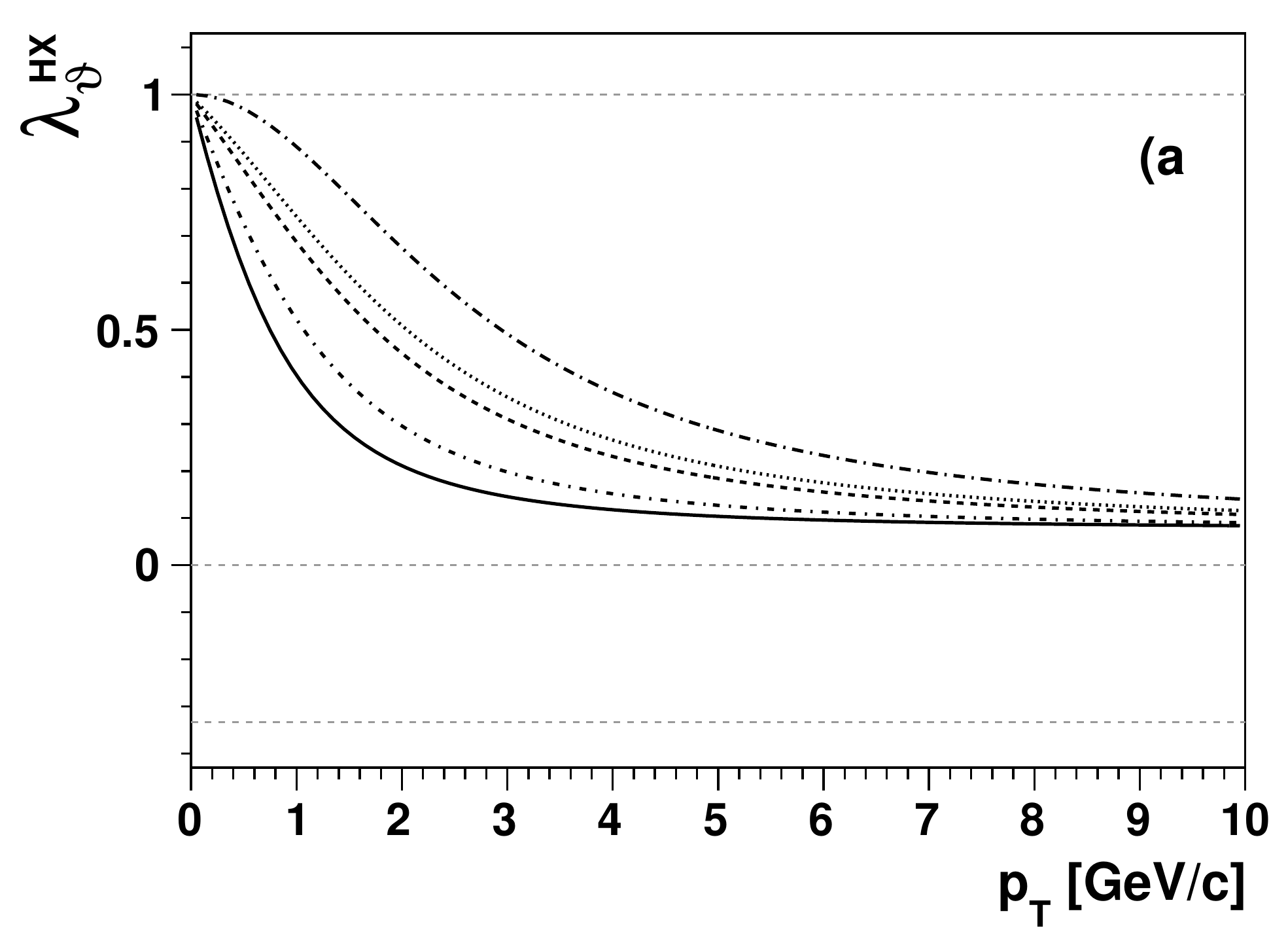}
\includegraphics{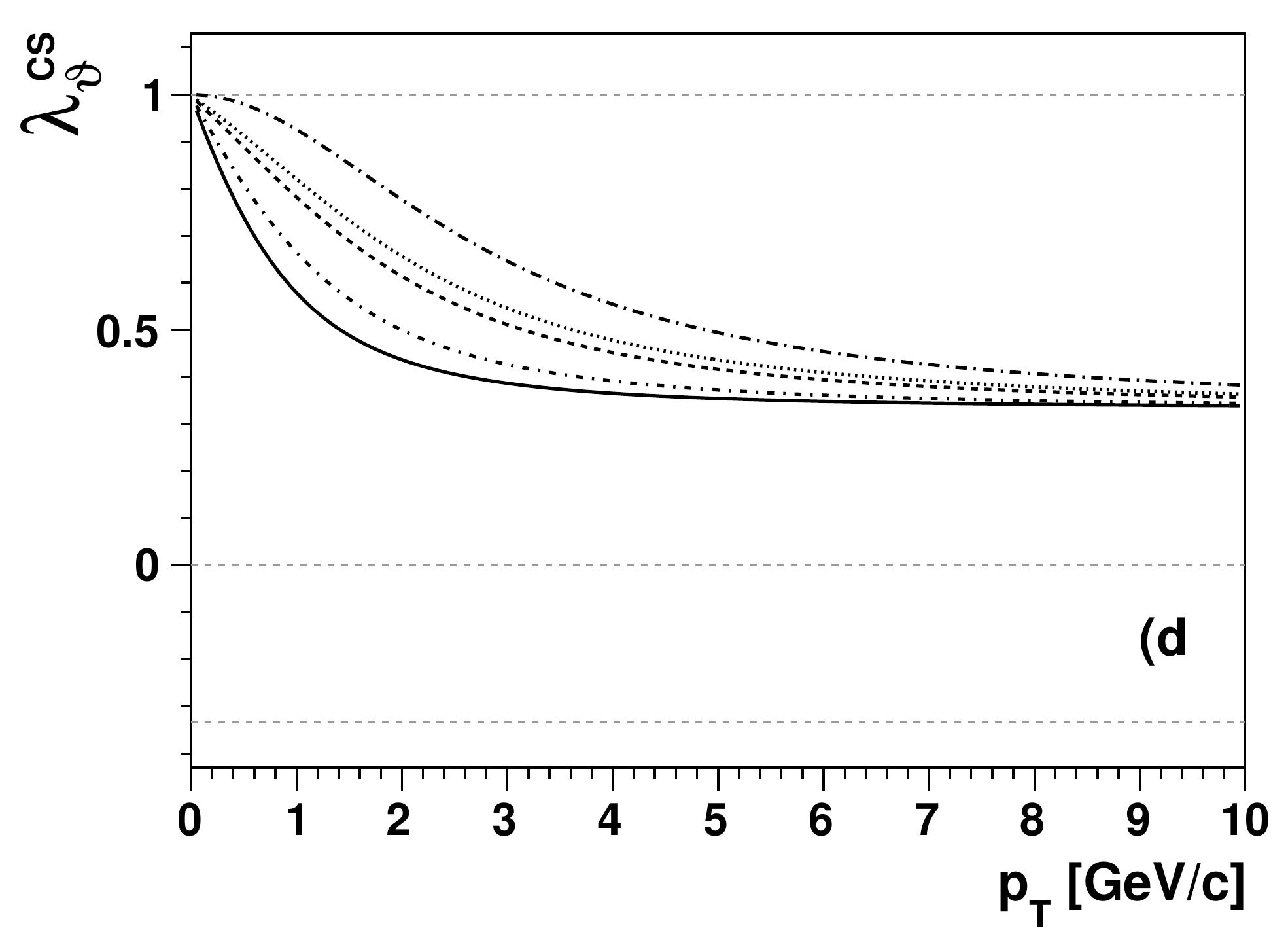}}
\resizebox{0.73\linewidth}{!}{%
\includegraphics{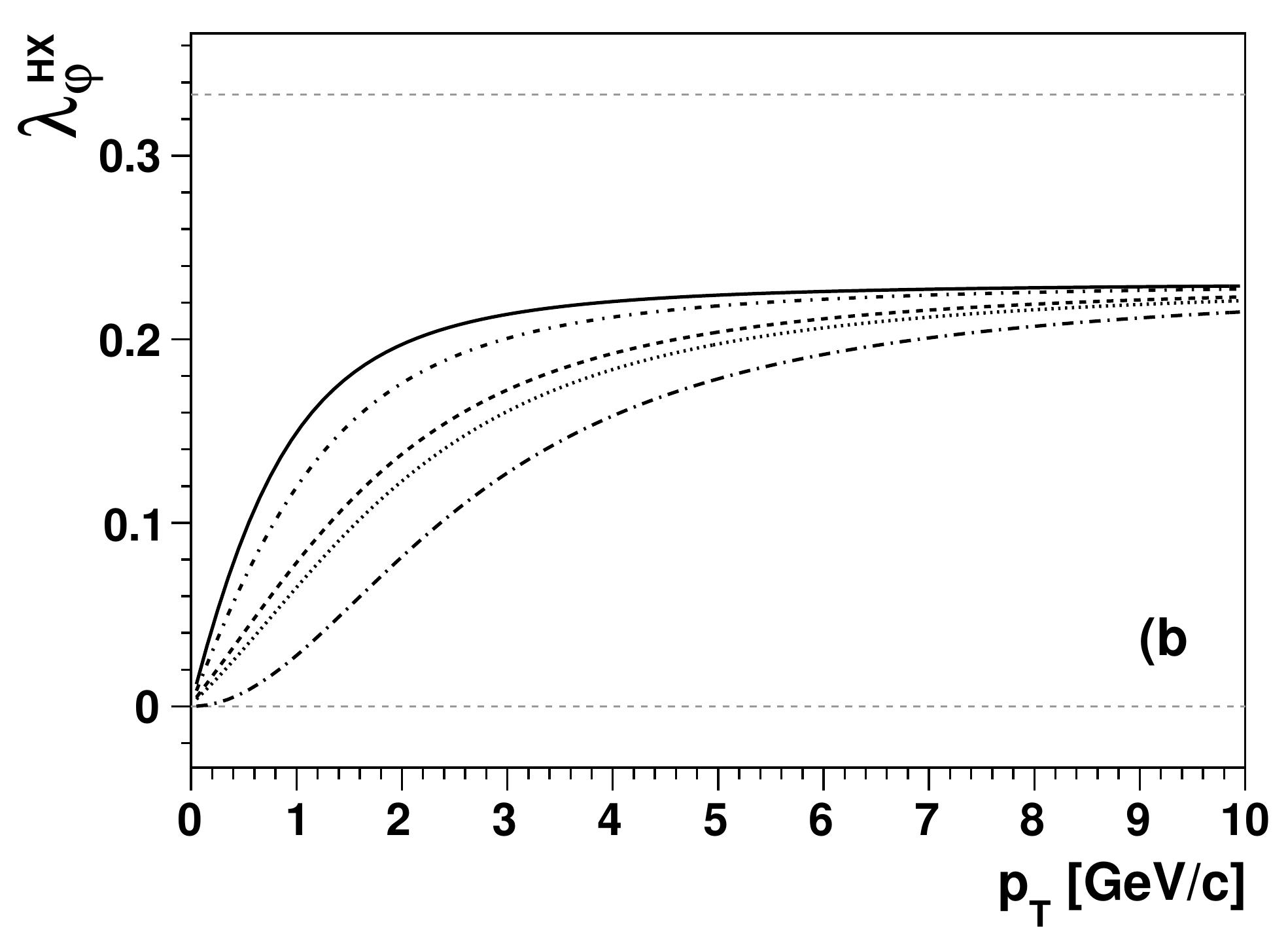}
\includegraphics{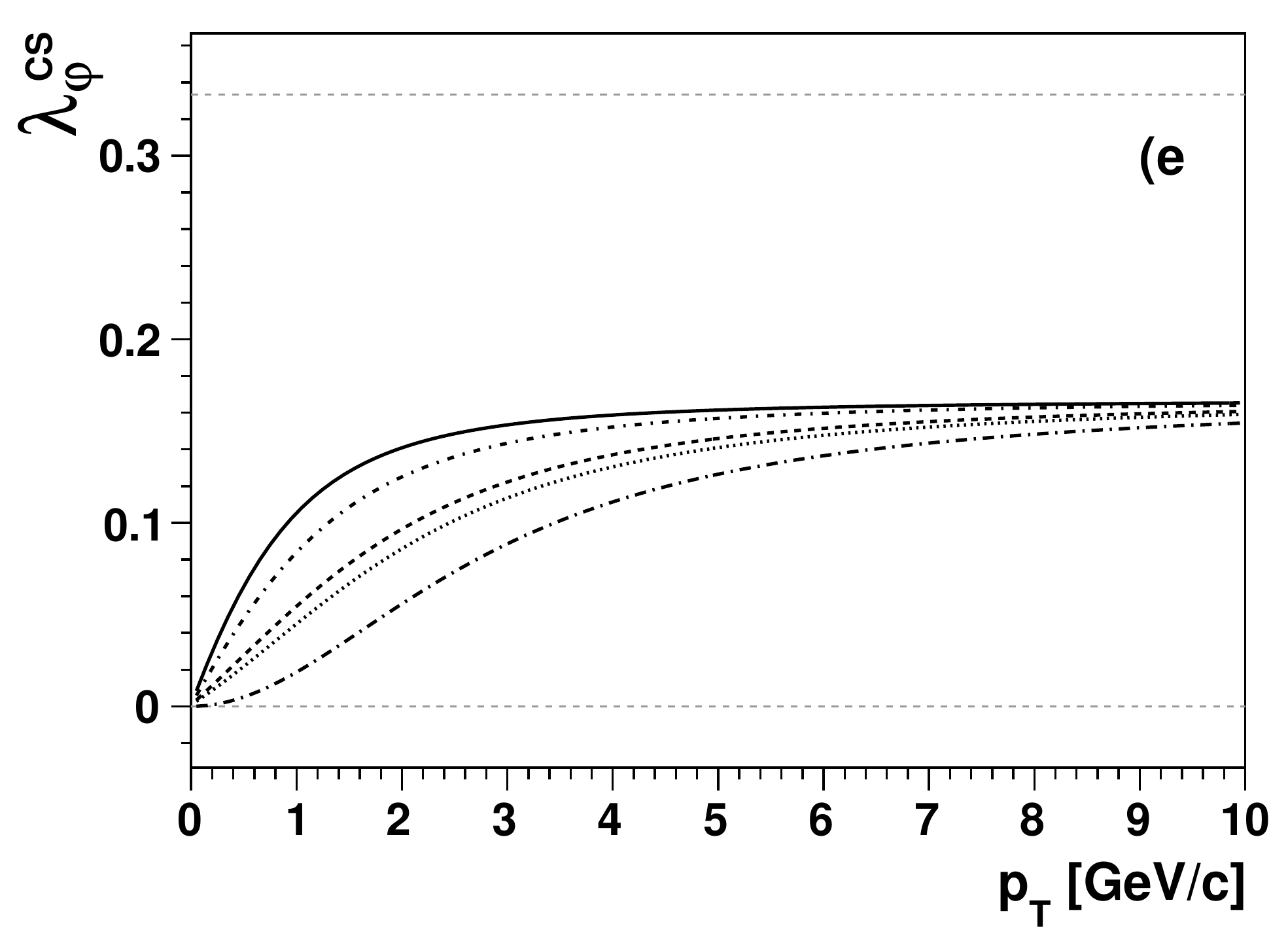}}
\resizebox{0.73\linewidth}{!}{%
\includegraphics{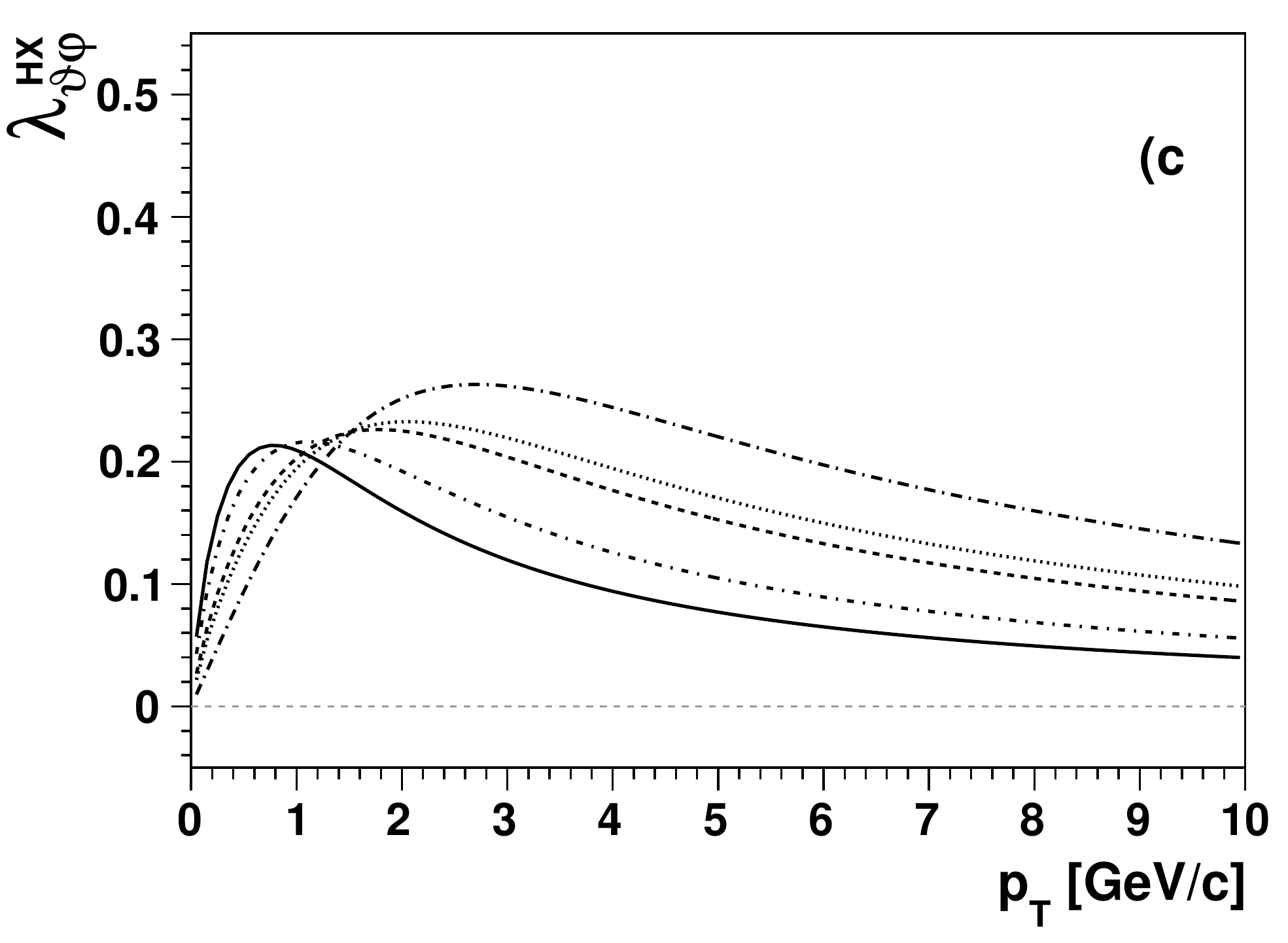}
\includegraphics{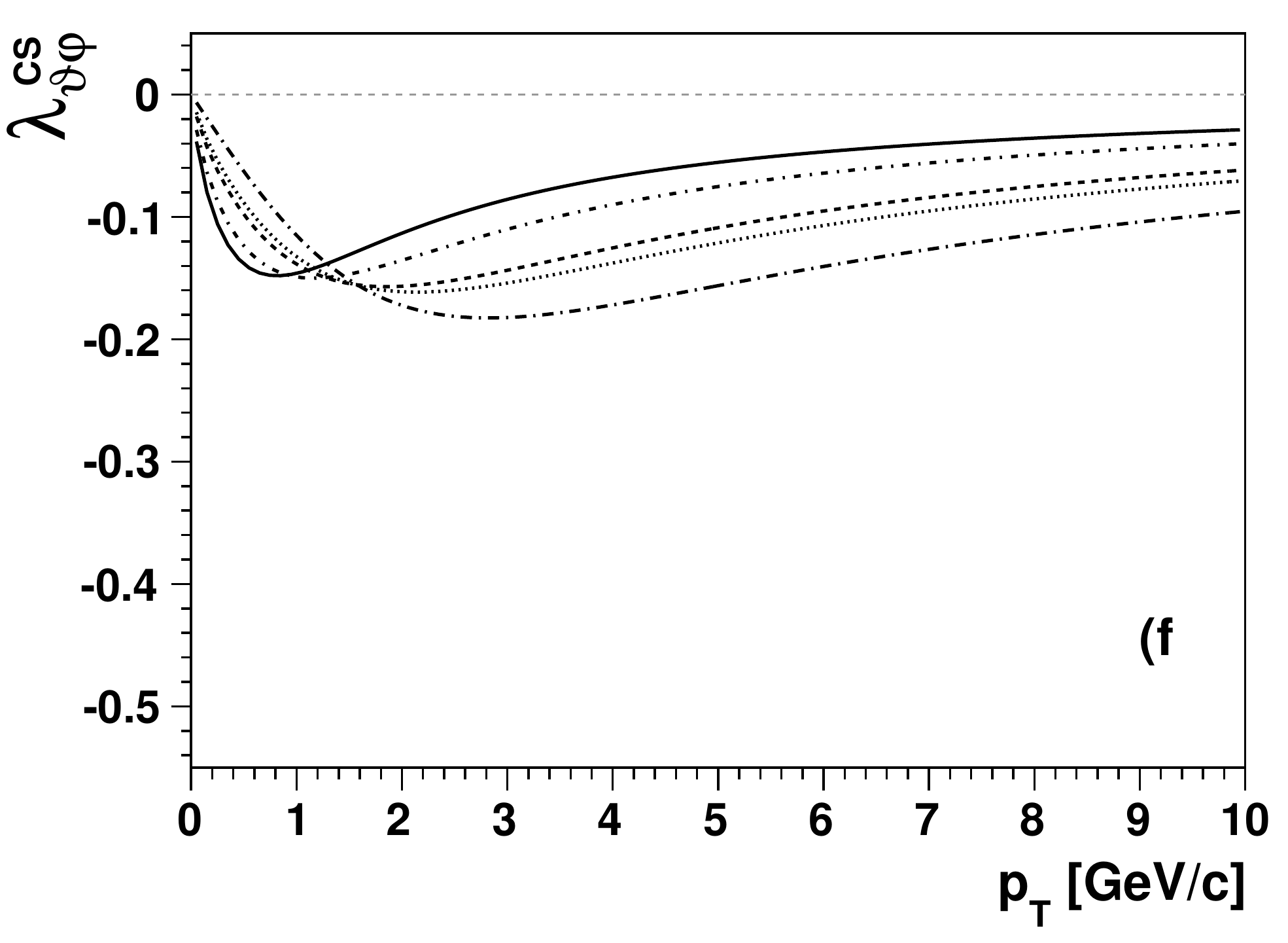}}
\caption{Kinematic dependence of the \jpsi\ decay angular distribution
  seen in the HX (a-b-c) and CS (d-e-f) frames, when $60\%$ ($40\%$)
  of the events have full transverse polarization in the CS (HX)
  frame. The curves represent measurements in different acceptance
  ranges, as detailed in Fig.~\ref{fig:kindep_lambda}.
  Corresponding figures for the \upsOneS\ case can be seen in
  Ref.~\cite{bib:ImprovedQQbarPol}.}
\label{fig:kindep_lambda_mix}
\end{figure*}
%%%%%%%%%%%%%%%%%%%%%%%%%%%%%%%%%%%%%%%%%%%%%%%%%%
This is shown in Fig.~\ref{fig:kindep_lambda_mix}, where we consider,
for illustration, that $60\%$ of the \jpsi\
%\upsAll\ 
events have natural
polarization $\lambda_\vartheta = +1$ in the CS frame while the
remaining fraction has $\lambda_\vartheta = +1$ in the HX frame.
Although the polarizations of the two event subsamples are intrinsically
independent of the production kinematics, in neither frame, CS or HX,
will measurements performed in different transverse and longitudinal
momenta windows find identical results.
However, in this case
%For instance, in the case illustrated in Fig.~\ref{fig:kindep_lambda_mix},
as well as in the simpler case of Fig.~\ref{fig:kindep_lambda}\,(a-b-c),
any arbitrary choice of the experimental observation frame will always
yield the value $\tilde{\lambda} = +1$, independently of kinematics.
This particular case, where all contributing processes are
transversely polarized, is formally equivalent to the Lam-Tung
relation~\cite{bib:LamTung}, as discussed in
Ref.~\cite{bib:LTGen}.  Analogously, the example represented in
Fig.~\ref{fig:kindep_lambda}\,(d-e-f), or any other case where all
polarizations are longitudinal, yields $\tilde{\lambda} = -1$.

The existence of frame-invariant parameters also provides a useful
instrument for experimental analyses. Checking, for example, that the
same value of an invariant quantity (Eq.~\ref{eq:invariants}) is
obtained, within systematic uncertainties, in two distinct
polarization frames is a non-trivial verification of the absence of
unaccounted systematic effects.
In fact, detector geometry and/or data selection constraints may
strongly polarize the reconstructed dilepton events.  Background
processes also affect the measured polarization, if not well
subtracted.
The spurious anisotropies induced by detector effects and background
do not obey the frame transformation rules characteristic of a
physical $J=1$ state.
If not well corrected and subtracted, these effects will distort the
shape of the measured decay distribution differently in different
polarization frames.  In particular, they will violate the
frame-independent relations between the angular parameters.
Any two physical polarization axes (defined in the rest frame of the
meson and belonging to the production plane) may be chosen to perform
these ``sanity tests''. The HX and CS frames are ideal choices at high
\pt, where they tend to be orthogonal to each other (in
Eq.~\ref{eq:delta2_HX_to_CS}, $\sin^2 \delta \rightarrow 1$ for
$p_\mathrm{T} \gg m$ and/or $p_\mathrm{T} \gg |p_\mathrm{L}|$).
At low \pt, where the difference between the two frames vanishes, any
of the two and its exact orthogonal may be used to maximize the
significance of the test.
Given that $\tilde{\lambda}$ is ``homogeneous'' to the anisotropy
parameters, the difference $\tilde{\lambda}^{({\rm B})} -
\tilde{\lambda}^{({\rm A})}$ between the results obtained in two
frames provides a direct evaluation of the level of systematic errors
not accounted in the analysis.

%%%%%%%%%%%%%%%%%%%%%%%%%%%%%%%%%%%%%%%%%%%%%%%%%%%%%%%%%%%%%%%%%%%%%%%%%%%%%%%%%%%
\section{Effect of intrinsic parton transverse momentum}
%\section{Effect of the parton transverse momentum}
\label{sec:kt}

In this section we describe how the geometry of the CS frame is
related to the kinematics of the production process. It can be
recognized from Eq.~\ref{eq:angle_CS_HX} that the vector
$\hat{z}_{\mathrm{CS}}$ indicates the direction of the
\emph{laboratory} $z$ axis (that is, the beam line) as seen in the
quarkonium rest frame.
In this frame, any length will be Lorentz contracted by a factor
$1/\gamma$ along the quarkonium boost direction, but not along the
transverse directions.
%a one-dimensional rod placed at
%rest along the beam line undergoes a contraction by $1/\gamma$ of its
%projection along the quarkonium boost direction, while the length of
%its transverse projection remains unchanged.
%
In the quarkonium rest frame (as well as in the laboratory)
the direction of the beam line coincides with the direction of the
relative motion of the colliding partons (``parton axis''), when their
transverse momenta are neglected (and exactly when averaging a large
sample of events).  This approximation affects
%We will now evaluate the importance of this latter approximation in 
the experimental determination of an angular
distribution naturally of the kind $1+ \lambda_\vartheta^{*} \cos^2
\vartheta^{*}$ with respect to the parton axis $z^{*}$. In the
following considerations we fix a coordinate system having the $z$
axis along the dilepton direction in the laboratory and the $xz$ plane
coinciding with the production plane. We then define the directions of
the beam axis and of the parton axis in the laboratory as,
respectively,
\begin{align}
\begin{split}
\hat{B} & =        ( \sin \Theta, 0, \cos \Theta ) \, , \\
\hat{B}^\prime & = ( \sin \Theta^\prime \cos \Phi^\prime, \sin \Theta^\prime
\sin \Phi^\prime, \cos \Theta^\prime ) \, , \label{eq:beam_partons_lab}
\end{split}
\end{align}
where $\Theta$ and $\Theta^\prime$ are the angles they form with respect to the
dilepton direction. The presence of the angle $\Phi^\prime$ denotes the fact
that, due to the intrinsic transverse momenta of the partons, the vector
$\hat{B}^\prime$ does not belong, in general, to the production plane. The
angle $\Delta$ between the two directions in the laboratory is given by
\begin{equation}
\cos \Delta = \sin \Theta \sin \Theta^\prime \cos \Phi^\prime + \cos \Theta
\cos \Theta^\prime \, . \label{eq:Delta_beam_partons}
\end{equation}
When boosted to the dilepton rest frame, the two vectors become
\begin{align}
\begin{split}
\hat{b} & =      \frac{( \sin \Theta, 0, \frac{1}{\gamma} \cos \Theta )}
{\sqrt{\sin^2 \Theta + \frac{1}{\gamma^2} \cos^2 \Theta }}   \, , \\
\hat{b}^{\prime} & = \frac{( \sin \Theta^\prime \cos \Phi^\prime, \sin
\Theta^\prime \sin \Phi^\prime, \frac{1}{\gamma} \cos \Theta^\prime
)}{\sqrt{\sin^2 \Theta^\prime + \frac{1}{\gamma^2} \cos^2 \Theta^\prime }}
\label{eq:beam_partons_dimuon}
\end{split}
\end{align}
and the cosine of the angle between them
\begin{equation}
\cos \zeta = \frac{\cos \Delta - \beta^2 \cos \Theta \cos \Theta^\prime }{
\sqrt{1 - \beta^2 \cos^2 \Theta} \sqrt{1 - \beta^2 \cos^2 \Theta^\prime} } \, .
\label{eq:zeta_beam_partons}
\end{equation}
The rotation by $\zeta$ from the parton axis to the beam line axis transforms
the polarization parameter $\lambda_\vartheta$ according to the following
expressions:
\begin{align}
\begin{split}
 \lambda^{\rm{CS}}_\vartheta  &  \; \simeq \;
\left(  1- \frac{3+\lambda_\vartheta^{*}}{2} \, \langle \sin^2\zeta \rangle
\right) \;
\lambda_\vartheta^{*} \, , \\
 \lambda^{\rm{CS}}_\varphi  & \; \simeq \;
 \lambda^{\rm{CS}}_{\vartheta \varphi}  \; \simeq \; 0 \, .
\label{eq:lambda_transf_parton_CS}
\end{split}
\end{align}
These transformations do not represent a simple rotation as
Eq.~\ref{eq:lambda_transf}. Indeed, the magnitude of the polar
anisotropy decreases, while no significant azimuthal anisotropy
arises. In fact, the rotation plane (formed by the parton and beam
lines) does not coincide with the experimentally defined production
plane. The angle between the two planes changes from one event to the
next, so that the azimuthal anisotropy deriving from the tilt between
the ``natural'' polarization axis and the experimental axis tends to
be smeared out in the integration over all events.
Since $\cos \Delta \simeq 1 - \frac{1}{2} \sin^2\Delta$ and
(approximately event-by-event, and exactly on average) $\cos
\Theta^\prime \simeq \cos \Theta$, from Eq.~\ref{eq:zeta_beam_partons}
we obtain
\begin{equation}
\langle \sin^2\zeta \rangle \; \simeq \; \frac{\langle \sin^2\Delta \rangle}{ 1
- \beta^2 \cos^2 \Theta } \; = \; \frac{E^2}{m_{\rm{T}}^2} \, \langle \sin^2\Delta
\rangle \, . \label{eq:sinzeta2_beam_partons}
\end{equation}
Denoting by $\vec{k}_{1,2}$, $\vec{k}_{1,2\rm{T}}$ and $E_{1,2}$
the total momenta, transverse momenta and energies of the two partons
in the laboratory, the laboratory angle $\Delta$ satisfies
\begin{equation}
\sin^2\Delta \; = \; \frac{( \vec{k}_{1\rm{T}} - \vec{k}_{2\rm{T}} )^2 }{ (
\vec{k}_{1} - \vec{k}_{2} )^2 } \; \simeq \; \frac{( \vec{k}_{1\rm{T}} -
\vec{k}_{2\rm{T}} )^2 }{ ( E_{1} + E_{2} )^2 } \label{eq:sinDelta2}
\end{equation}
and, on average,
\begin{equation}
\langle \sin^2\Delta \rangle \; \simeq \; \frac{ 2 \langle \vec{k}_{\rm{T}}^2
\rangle }{ ( E_{1} + E_{2} )^2 } \, , \label{eq:sinDelta2avg}
\end{equation}
where we have defined the average parton squared transverse momentum
as $\langle \vec{k}_{\rm{T}}^2 \rangle = (\langle \vec{k}_{1\rm{T}}^2
\rangle +\langle \vec{k}_{2\rm{T}}^2 \rangle)/2$.

Considering now the specific case of Drell-Yan production at low \pt,
we can assume an approximate equality between total parton energy and
dilepton energy, $E_{1} + E_{2} \simeq E$, and, moreover, $\langle
m_{\rm{T}}^2 \rangle \simeq m^2 + 2 \langle
\vec{k}_{\rm{T}}^2\rangle$. Combining
Eqs.~\ref{eq:lambda_transf_parton_CS} (with $\lambda_\vartheta^{*} =
+1$), \ref{eq:sinzeta2_beam_partons} and \ref{eq:sinDelta2avg}, we
find that the measurement of the polarization of low-\pt\ Drell-Yan
dileptons provides an estimate of the ``effective'' parton transverse
momentum:
\begin{equation}
\langle \vec{k}_{\rm{T}}^2 \rangle \; \simeq \; \frac{m^2}{2} \;
\frac{1-\lambda_\vartheta^{\rm{CS}}}{1+\lambda_\vartheta^{\rm{CS}}} \, .
\label{eq:kT2_lambdatheta}
\end{equation}
The average Drell-Yan polarization $\lambda_\vartheta^{\rm{CS}} =
1.008 \pm 0.026$ measured by E866~\cite{bib:e866_Ups}, in
proton-copper collisions for $\langle m \rangle \simeq 10$~GeV$/c^2$
and $p_{\rm{T}} < 4$~GeV$/c$, implies, therefore, that
\begin{align}
\begin{split}
\langle \vec{k}_{\rm{T}}^2 \rangle \; & < \; 0.5 \; {\rm GeV}^2/c^2
\quad {\rm at~ 68\% ~ C.L.\ and}\\ 
\langle \vec{k}_{\rm{T}}^2 \rangle \; & < \; 1.0 \; {\rm GeV}^2/c^2 
\quad {\rm at~ 95\% ~ C.L.\, .}
\label{eq:kT2_limit_E866}
\end{split}
\end{align}
Tighter limits could be derived from precise low-mass measurements,
given that the polarization smearing effects are essentially
proportional to $m^{-2}$.  Unfortunately, the existing (pion-induced)
measurements~\cite{bib:NA10,bib:E615}, though very precise and
extending down to $4$~GeV$/c^2$, present large azimuthal anisotropies
of dubious interpretation and are scarcely suitable for this purpose.

We can now estimate the maximum magnitude of the smearing effects that
can be foreseen for the observable quarkonium polarization when the
natural polarization axis is the parton axis. Combining again
Eqs.~\ref{eq:lambda_transf_parton_CS}, \ref{eq:sinzeta2_beam_partons}
and \ref{eq:sinDelta2avg}, this time with $E_{1} + E_{2} \geq E$, we
find that the magnitude of the polarization is reduced by the fraction
\begin{equation}
\left| \frac{\lambda_\vartheta^{\rm{CS}} -
\lambda_\vartheta^{*}}{\lambda_\vartheta^{*}} \right| \; \lesssim \; \frac{(3 +
\lambda_\vartheta^{*}) \langle \vec{k}_{\rm{T}}^2 \rangle}{m^2 + p_{\rm{T}}^2}
\, . \label{eq:lambdatheta_kT_smearing}
\end{equation}
For example, it cannot be excluded, considering the limit in
Eq.~\ref{eq:kT2_limit_E866}, that a fully transverse natural
polarization of the \jpsi\ along the parton axis is reduced by as much
as $30\%$, for $p_{\rm{T}} = 2$~GeV$/c$, when observed in the CS
frame.
This smearing effect should be one order of magnitude smaller for
\jpsi\ mesons of $p_{\rm{T}} = 10$~GeV$/c$. On the other hand, given
the strong dependence of the effect on the dilepton mass, bottomonium
polarization measurements are practically insensitive to the parton
transverse momentum even at low $p_{\rm{T}}$. This prediction is
consistent with the already quoted E866 results, showing
$\Upsilon(2S+3S)$ polarizations in the CS frame always compatible with
$+1$, within a $\sim 15\%$ uncertainty, in four $p_{\rm{T}}$ bins
between $0$ and $4$~GeV$/c$.

\section{A few concrete examples}
\label{sec:examples}

We conclude with some examples of measurements illustrating concepts
described in the previous sections.

\begin{figure*}[ht]
\centering
\resizebox{0.95\linewidth}{!}{%
\includegraphics{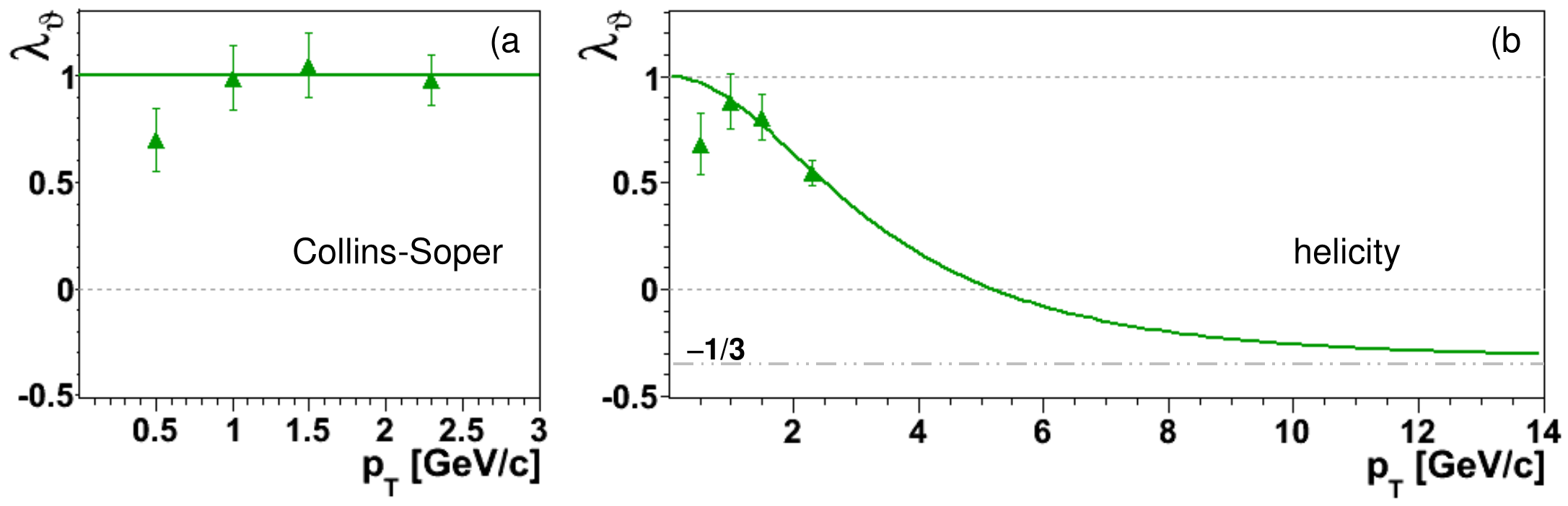}}
\caption{The E866 measurement of the $\Upsilon(2S+3S)$ polarization in
  the CS (a) and a guess of how it would be observed in the HX frame,
  extrapolating to higher \pt\ (b).} \label{fig:E866}
\end{figure*}

We have already referred to the E866 measurement of a full transverse
$\Upsilon(2S+3S)$ polarization in the CS frame. The result is
represented in Fig.~\ref{fig:E866}\,a, as a function of $p_{\rm{T}}$. A
similarly constant behaviour, consistent with a Drell-Yan-like
polarization, has been measured by this experiment as a function of
$x_{\rm{F}}$, in the range $[0,0.5]$, confirming that the adoption of
the CS frame is, in this case, an optimal choice.
It is true that, in special kinematic conditions, the transverse
polarization observed in one frame could, in reality, be the
reflection of a natural longitudinal polarization in another frame, as
shown in Fig.~\ref{fig:kindep_lambda}\,d.  However, a maximal
polarization \emph{independent of the production kinematics} in the CS
frame must directly reflect the spin configuration of the interacting
partons (as is well known to be the case in Drell-Yan production, a
paradigmatic example of natural transverse polarization in the CS
frame).

To better illustrate the importance of an optimal choice of the
reference frame, we will now consider what the E866 experiment would
have measured, had the analysis been made with a different choice.
As a reasonable approximation, we assume that the azimuthal
distribution is exactly isotropic in the CS frame.  The polar
anisotropy that would be observed in the HX frame is shown in
Fig.~\ref{fig:E866}\,b, where the curve includes an extrapolation to
higher $p_{\rm{T}}$ assuming that in the CS frame the distribution
continues to have the shape $1 + \cos^2 \vartheta$. A measurement
performed in the HX frame would show, quite misleadingly, a
polarization changing from fully transverse to partially
longitudinal. The strong signature of ``natural'' transverse
polarization evidenced by the data in the CS frame would become
unrecognizable, although it could, in principle, be reconstructed back
if (and only if) the azimuthal anisotropy were also measured.
%
%%%%%%%%%%%%%%%%%%%%%%%%%%%%%%%%%%%%%%%%%%%%
%%%%%%%%%%%%%%%%%%%%%%%%%%%%%%%%%%%%%%%%%%%%%%%%%%
%

\begin{figure}[h]
\centering
\resizebox{0.5\linewidth}{!}{%
\includegraphics{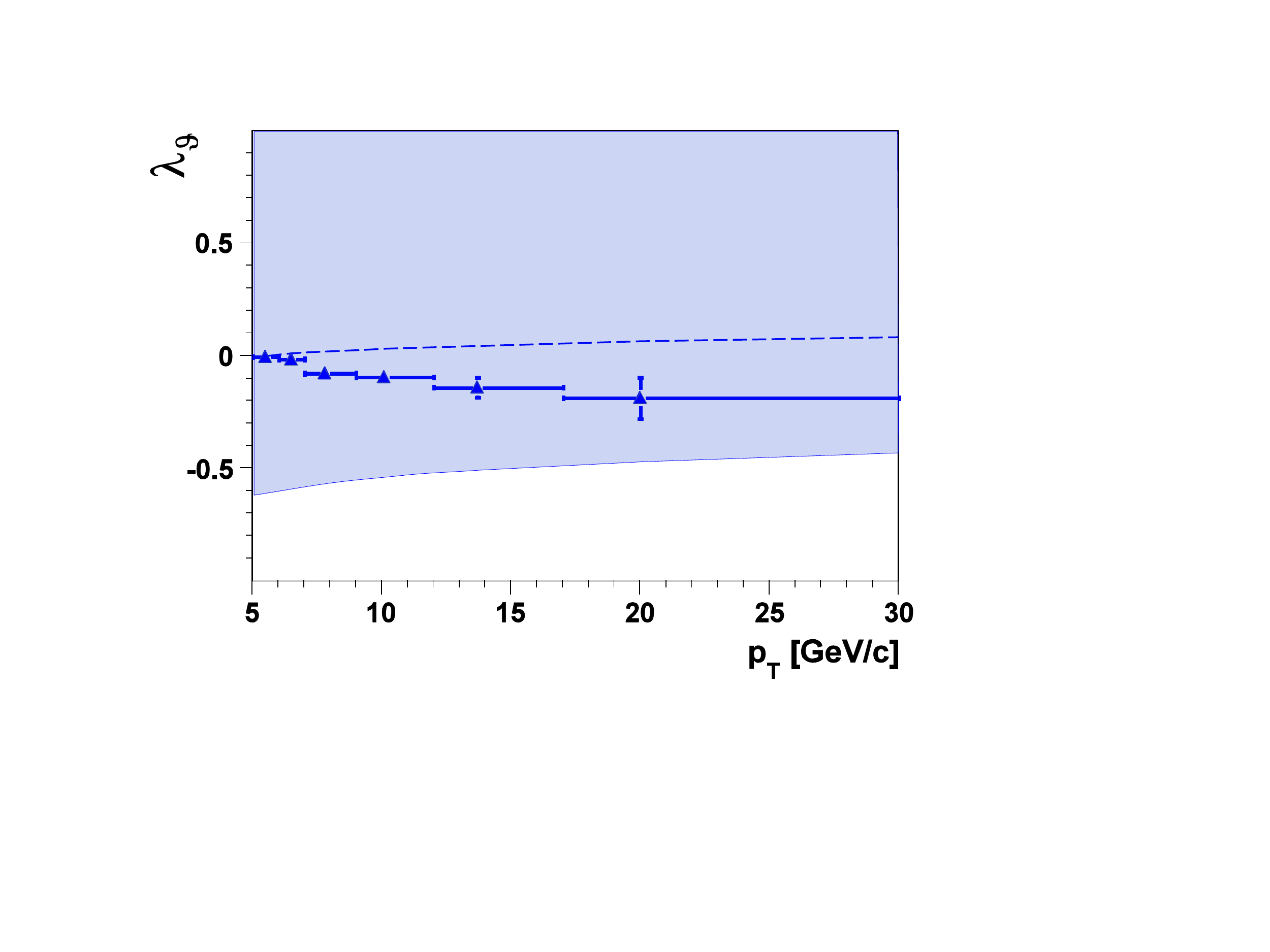}}
\caption{The CDF \jpsi\ polarization measurement in the helicity frame (data
points) and the range for the corresponding polarization in the CS, allowing
for all possible values of the azimuthal anisotropy (shaded band). The dashed
line is the CS polarization for $\lambda_\varphi^{\rm{HX}}=0$.}
\label{fig:CDF_CS}
\end{figure}

Seeing how the curve in Fig.~\ref{fig:E866}\,b qualitatively
resembles the pattern measured by CDF for the \jpsi, it is natural to
wonder how that measurement, made in the HX frame, would look like in
the CS frame. Unfortunately, in this case the measurement itself, a
slight longitudinal polarization, does not suggest any educated guess
on what we could assume for the unmeasured azimuthal anisotropy.  For
example, as shown in Fig.~\ref{fig:CDF_CS}, if the distribution in
the HX frame were azimuthally isotropic, the measured polarization
would correspond to a practically undetectable polarization in the CS
frame (dashed line).
%
%%%%%%%%%%%%%%%%%%%%%%%%%%%%%%%%%%%%%%%%%%%%
%%%%%%%%%%%%%%%%%%%%%%%%%%%%%%%%%%%%%%%%%%%%%%%%%%
%
However, if we take into account all
physically possible values of the azimuthal anisotropy, as allowed by
the triangular relation represented in Fig.~\ref{fig:triangles}, we
can only derive a broad spectrum of possible CS polarizations,
approximately included between $-0.5$ and $+1$ (shaded band).
This example shows how a measurement reporting only the polar
anisotropy is amenable to several interpretations in fundamental
terms, often corresponding to drastically different physical cases.
One possible hypothesis would be that all processes are naturally
polarized in the HX frame and that transverse and longitudinal
polarizations are superimposed in proportions varying from
approximately $2/3$ transverse and $1/3$ longitudinal at $p_{\rm T} =
5$~GeV$/c$ ($\lambda_\vartheta \simeq 0$) to around 60\%\,/\,40\% at
$p_{\rm T} = 20$~GeV$/c$ ($\lambda_\vartheta \simeq -0.2$).
In this case, no azimuthal anisotropy should be observed in the HX
frame.
Alternatively, we can consider a scenario where the observed
\emph{slightly longitudinal} HX polarization is actually the result of
a mixture of two processes, both producing \jpsi\ mesons with
\emph{fully transverse} polarizations, but one in the HX frame and the
other in the CS frame.
%
%
%%%%%%%%%%%%%%%%%%%%%%%%%%%%%%%%%%%%%%%%%%%%
\begin{figure}[ht]
\centering
\resizebox{0.48\linewidth}{!}{%
\includegraphics{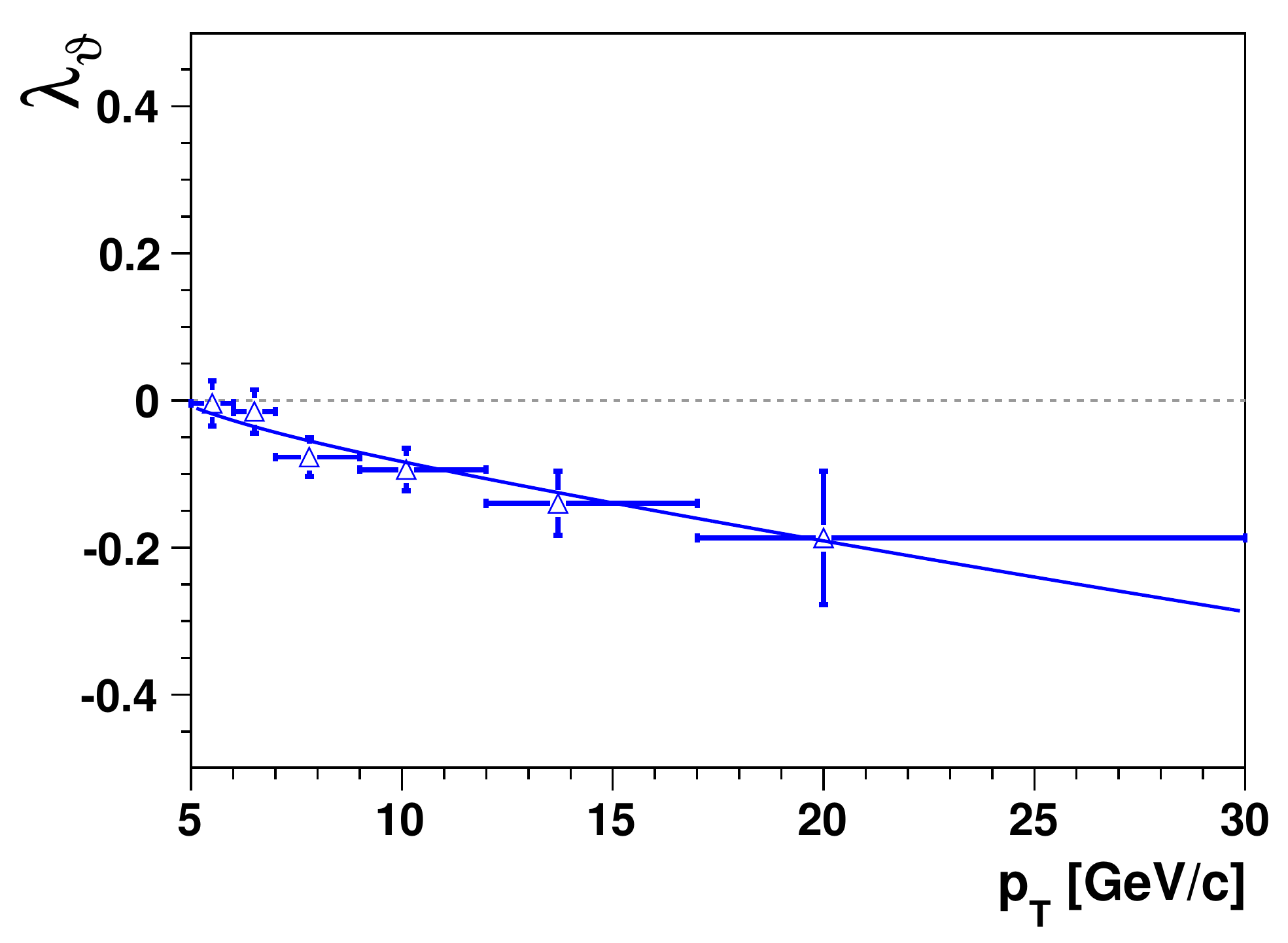}}
\resizebox{0.48\linewidth}{!}{%
\includegraphics{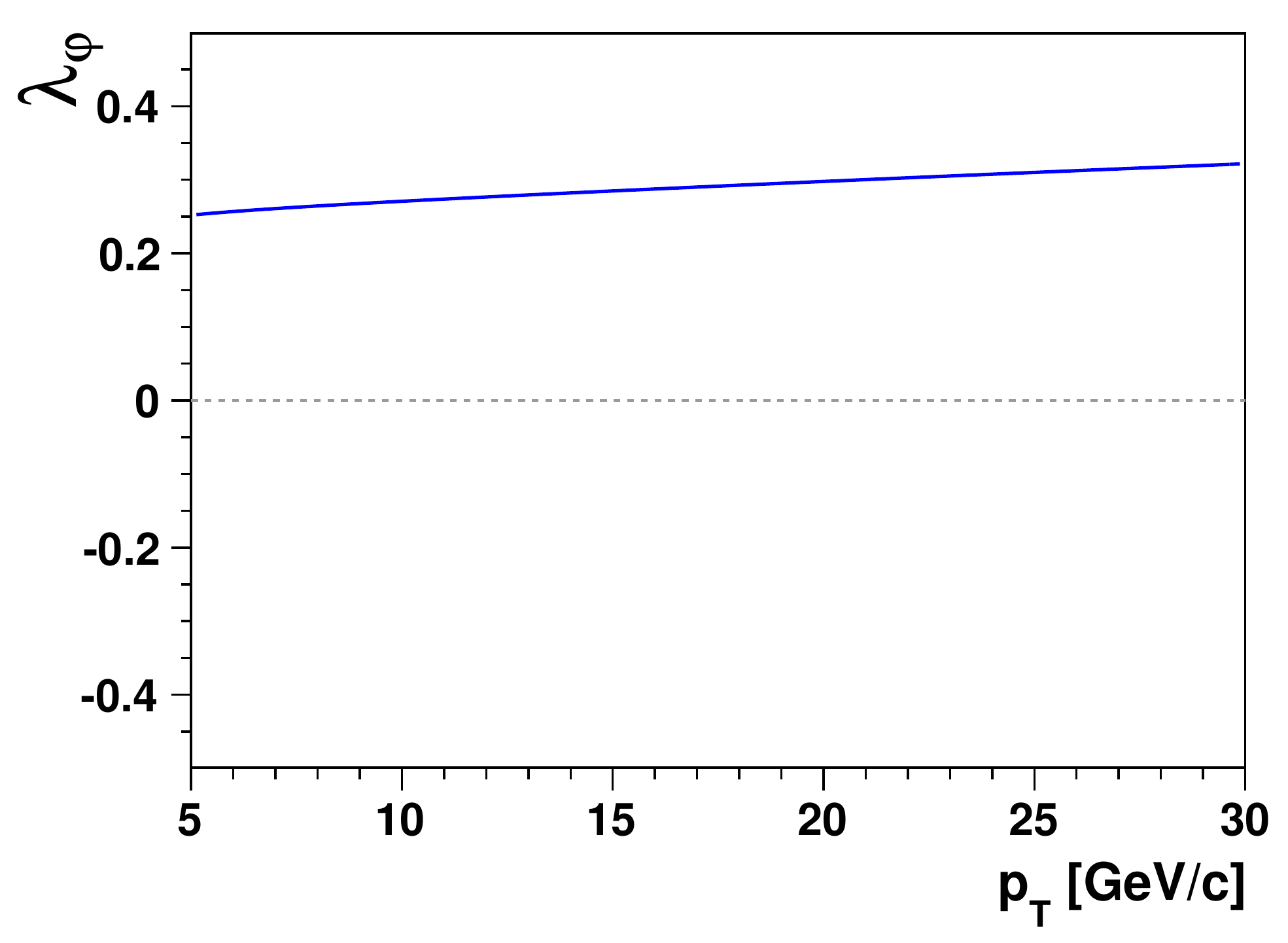}}
\caption{The \pt\ dependence of the angular parameters
  $\lambda_\vartheta$ (left) and $\lambda_\varphi$ (right) as would be
%%%  $\lambda_\vartheta$ (top) and $\lambda_\varphi$ (bottom) as would be
  measured by CDF in the HX frame, according to a scenario where the
  \jpsi's have always full transverse polarization, either in the CS
  frame or in the HX frame, with a suitable \pt-dependent proportion
  between the two event samples. The data points represent the CDF
  measurement.}
\label{fig:CDF_transverse}
\end{figure}
%%%%%%%%%%%%%%%%%%%%%%%%%%%%%%%%%%%%%%%%%%%%%%%%%%
%%%Figure~\ref{fig:CDF_transverse}\,(top) shows that this scenario is
Figure~\ref{fig:CDF_transverse}\,(left) shows that this scenario is
perfectly compatible with the CDF $\lambda_\vartheta$ measurement if
the proportion $f_{\rm HX}/(f_{\rm HX} + f_{\rm CS})$ between the two
sub-processes is assumed to vary linearly between $30\%$ at 
$p_{\rm T} = 5$~GeV$/c$ and $15\%$ at $p_{\rm T} = 20$~GeV$/c$.
The difference with respect to the previous hypothesis is that now we
would measure a significant azimuthal anisotropy, $\lambda_\varphi
\simeq 0.3$, 
%%%as shown in Fig.~\ref{fig:CDF_transverse}\,(bottom).
as shown in Fig.~\ref{fig:CDF_transverse}\,(right). 
As an attempt to reconcile low-\pt\ measurements with collider
data, Ref.~\cite{bib:pol} described one further conjecture, in
which the polarization arises naturally in the CS frame, and becomes
increasingly transverse with increasing total \jpsi\ momentum. Again,
a direct measurement of $\lambda_\varphi$ (which, in this case, should
be zero in the CS frame but positive and increasing in the HX frame)
would easily clarify the situation.

%
%%%%%%%%%%%%%%%%%%%%%%%%%%%%%%%%%%%%%%%%%%%%
\begin{figure}[ht]
\centering
\resizebox{0.45\linewidth}{!}{%
\includegraphics{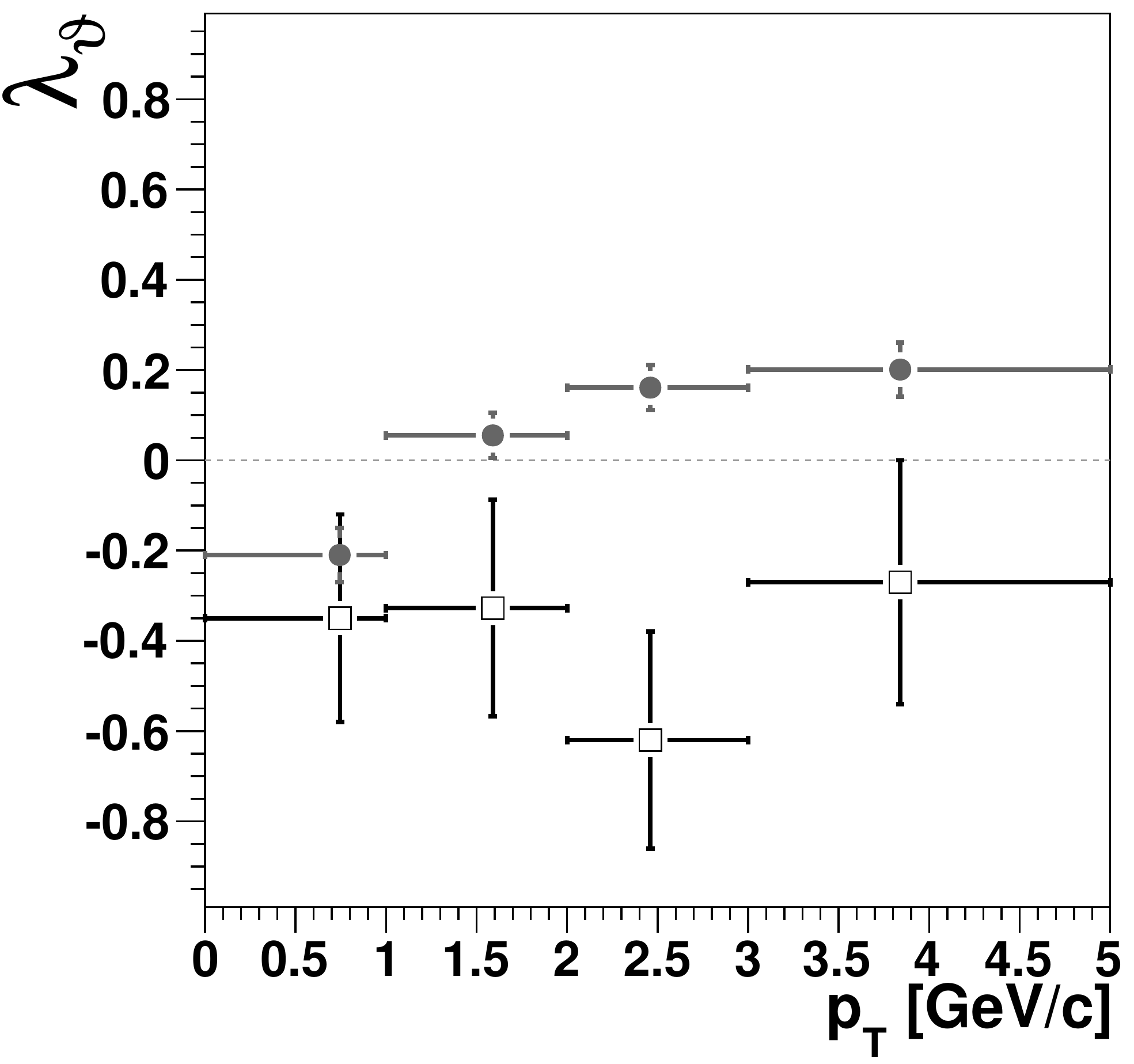}}
\resizebox{0.45\linewidth}{!}{%
\includegraphics{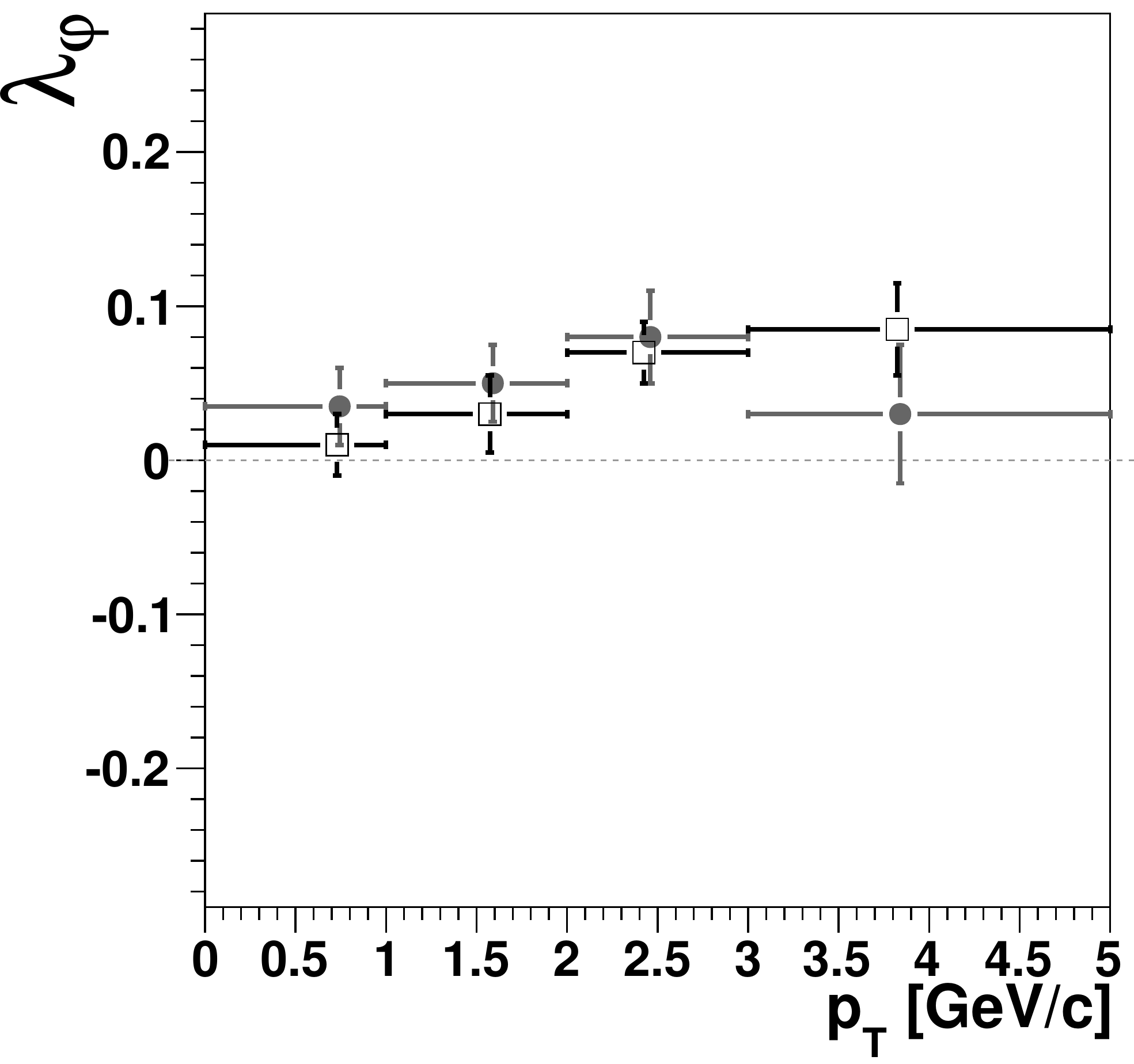}}
\caption{Example of a ``gedankenexperiment'' where the \jpsi\
  polarization measurements in the CS and HX frames (empty and filled
  symbols, respectively) would be inconsistent with each other.}
\label{fig:NA60}
\end{figure}
%%%%%%%%%%%%%%%%%%%%%%%%%%%%%%%%%%%%%%%%%%%%%%%%%%
%

We finish this section by illustrating the application of the
frame-independent formalism as a tool to estimate residual systematic
uncertainties in experimental data analyses. Figure~\ref{fig:NA60}
shows a putative set of \jpsi\ polarization measurements performed in
the CS and HX frames, versus \pt.  While the $\lambda_\vartheta$
values seem to change significantly from one frame to the other, the
two $\lambda_\varphi$ patterns are very similar.
This observation should alert to a possible experimental
artifact in the data analysis.  We can evaluate the significance of
the apparent contradiction by calculating the frame-invariant
$\tilde{\lambda}$ variable in each of the two frames.  For the case
illustrated in Fig.~\ref{fig:NA60}, averaging the four represented
\pt\ bins, we see that $\tilde{\lambda}$ in the HX frame is larger
than in the CS frame by 0.5 (a rather large value, considering that
the decay parameters are bound between $-1$ and $+1$).  In other
words, an experiment obtaining such measurements would learn from this
simple exercise that its determination of the decay parameters must be
biased by systematic errors of roughly this magnitude.
Given the puzzles and contradictions existing in the published
experimental results, as recalled in Section~\ref{sec:intro}, the use
of a frame-invariant approach to perform self-consistency checks,
which can expose unaccounted systematic effects due to detector
limitations and analysis biases, constitutes a non-trivial
complementary aspect of the new approach proposed in this paper for
quarkonium polarization measurements.

\section{Summary and conclusions}

Motivated by several puzzles affecting the existing measurements of
quarkonium polarization, we present in this paper a set of proposals
which should improve the experimental determination of the \jpsi\ and
\upsAll\ polarizations.  They are summarized in the next paragraphs.

Measurements and calculations of vector quarkonium polarization
should provide results for the full dilepton decay angular
distribution (a three-parameter function) and not only for the polar
anisotropy parameter. Only in this way can the measurements and
calculations represent unambiguous determinations of the average
angular momentum composition of the produced quarkonium state in terms
of the three base eigenstates, with $J_z = +1, 0, -1$.

It is advisable to perform the experimental analyses in at least
two different polarization frames. In fact, the self-evidence of
certain signature polarization cases (e.g.\ a full polarization with
respect to a specific axis) can be spoiled by an unfortunate choice of
the reference frame, which can lead to artificial (``extrinsic'')
dependencies of the results on the kinematics and on the experimental
acceptance.

The measured dependence of the polarization on the production
kinematics is necessarily influenced by the differential experimental
acceptance, i.e.\ by the kinematic distribution of the population of
the accepted events.  This problem, which is not solved by acceptance
corrections, can be minimized by providing the results in narrow cells
in quarkonium rapidity and transverse momentum.  Theoretical
calculations should be provided as event-level inputs to Monte Carlo
generators which can be tailored to the specific performance
capabilities of each experiment.

The decay angular distribution can be characterized by a
frame-independent quantity, such as $\tilde{\lambda}$, calculable in
terms of the polar and azimuthal anisotropy parameters.  The existence
of such frame-invariant quantities can be used during the data
analysis phase to perform self-consistency checks that can expose
previously unaccounted biases, caused, for instance, by the detector
limitations or by the event selection criteria.

Besides providing a much needed control over systematic experimental
biases, the variable $\tilde{\lambda}$ also provides relevant physical
information: it characterizes the \emph{shape} of the angular
distribution, reflecting ``intrinsic'' spin-alignment properties of
the decaying state, irrespectively of the specific geometrical
framework chosen by the observer.
For instance, we obtain $\tilde{\lambda}=+1$ for the shapes shown in
Fig.~\ref{fig:natpols}~(a) and (c), and $\tilde{\lambda}=-1$ for the
shapes shown in the panels (b) and (d).
A very important advantage of re-expressing the frame-dependent polar
and azimuthal anisotropies in terms of a frame-invariant quantity is
the exact cancellation of extrinsic dependencies on kinematics and
acceptances, enabling more robust comparisons with other experiments
and with theory.
The calculation of $\tilde{\lambda}$ requires, anyhow, the
determination of the full decay distribution in a chosen reference
frame and, obviously, does not replace this standard procedure.
Moreover, the three frame-dependent parameters ($\lambda_{\vartheta}$,
$\lambda_{\varphi}$ and $\lambda_{\vartheta\varphi}$) can provide
information on the \emph{direction} of the spin-alignment of the
decaying particle (when this direction is univocally defined) and,
therefore, on the topological properties of the dominant production
mechanism.  For instance, the measurement of a full transverse
polarization in the CS frame represents a direct observation of the
spin alignments of the interacting partons, as we know from Drell-Yan
production.
On the other hand, in the presence of a superposition of production
processes with polarizations along different axes, measuring the
frame-dependent anisotropies will not provide, in general, much
information on the polarizations involved or on their natural
alignment directions, while the value of $\tilde{\lambda}$ will
immediately tell us if the processes involved have a predominantly
transverse or longitudinal nature.

Stripped-down analyses which only measure the polar
anisotropy in a single reference frame, as often done in past
experiments, give more information about the frame selected by the
analyst (``is the adopted quantization direction an optimal choice?'')
than about the physical properties of the produced quarkonium (``along
which direction is the spin aligned, on average?'').
For example, a natural longitudinal polarization will give any desired
$\lambda_\vartheta$ value, from $-1$ to $+1$, if observed from a
suitably chosen reference frame.  Lack of statistics is not a reason
to ``reduce the number of free parameters'' if the resulting
measurements become ambiguous.  The forthcoming measurements of
quarkonium polarization in proton-proton collisions at the LHC have
the potential of providing a very important step forward in our
understanding of quarkonium production, if the experiments adopt a
more robust analysis framework, incorporating the ideas presented in
this paper.

%%%\begin{acknowledgement}

  P.F., J.S.\ and H.K.W.\ acknowledge support from Funda\c{c}\~ao para
  a Ci\^encia e a Tecnologia, Portugal, under contracts SFRH
  /BPD/42343/2007, CERN/FP/109343/2009 and SFRH/BPD/42138/2007.

%%%\end{acknowledgement}

%%%\vfill
%%%\pagebreak

%%%%%%%%%%%%%%%%%%%%%%%%%%%%%%%%%%%%%%%%%%%%%%%%%%%%%


\begin{thebibliography}{99}
\parskip=5pt \parsep=0.pt \itemsep=0.pt

\bibitem{bib:YellowRep-QWG} N.~Brambilla \emph{et al.} (QWG Coll.),
  CERN Yellow Report 2005-005, hep-ph/0412158, and references therein.

\bibitem{bib:cdf1-psis} F.~Abe \emph{et al.} (CDF Coll.),
  Phys. Rev. Lett. \textbf{79}, 572 (1997).

\bibitem{bib:NRQCD} G.T.~Bodwin, E.~Braaten and G.P.~Lepage,
  Phys. Rev. \textbf{D 51}, 1125 (1995); Phys. Rev. \textbf{D 55},
  5853E (1997).

\bibitem{bib:lansberg-HP08} J.P. Lansberg, Eur. Phys. J. \textbf{C 61}
  (2009) 693.

\bibitem{bib:BK} M. Beneke and M. Kr\"amer,
  Phys. Rev. \textbf{D 55}, 5269 (1997).

\bibitem{bib:Lei} A.K. Leibovich, 
  Phys. Rev. \textbf{D 56}, 4412 (1997).

\bibitem{bib:BKL} E. Braaten, B.A. Kniehl and J. Lee,
  Phys. Rev. \textbf{D 62}, 094005 (2000).

\bibitem{bib:CDFpol2} A. Abulencia \emph{et al.} (CDF Coll.),
  Phys. Rev. Lett. \textbf{99}, 132001 (2007).

\bibitem{bib:feeddown} P.~Faccioli, C. Louren\c{c}o, J. Seixas, and
  H.K. W\"ohri, J. High Energy Phys. \textbf{10}, 004 (2008).

\bibitem{bib:pol} P. Faccioli, C. Louren\c{c}o, J. Seixas and
  H.K. W\"ohri, Phys. Rev. Lett. \textbf{102}, 151802 (2009).

\bibitem{bib:upsCDF} D. Acosta \emph{et al.} (CDF Coll.),
  Phys. Rev. Lett. \textbf{88}, 161802 (2002).
  %CDF Public Note 9966 (2009).

\bibitem{bib:upsD0} V.M. Abazov \emph{et al.} (D0 Coll.),
  Phys. Rev. Lett. \textbf{101}, 182004 (2008).

\bibitem{bib:e866_Ups} C.N. Brown \emph{et al.} (E866 Coll.),
  Phys. Rev. Lett. \textbf{86}, 2529 (2001).

\bibitem{bib:CDFpol1} T. Affolder \emph{et al.} (CDF Coll.),
  Phys. Rev. Lett. \textbf{85}, 2886 (2000).

\bibitem{bib:LamTung} C.S. Lam and W.K. Tung, Phys. Rev. \textbf{D
    18}, 2447 (1978).
%  \plb{80}{1979}{228};  \prd{21}{1980}{2712}.

\bibitem{bib:chic_angulardistr_CLEO} M. Artuso \emph{et al.} (CLEO
  Coll.), Phys. Rev. \textbf{D 80}, 112003 (2009).

\bibitem{bib:chic_angulardistr_Fermilab} T.A. Armstrong \emph{et al.}
  (E760 Coll.), Phys. Rev. \textbf{D 48}, 3037 (1993); M. Ambrogiani
  \emph{et al.} (E835 Coll.), Phys. Rev. \textbf{D 65}, 052002 (2002).

\bibitem{bib:psip_E835} M. Ambrogiani \emph{et al.} (E835 Coll.),
  Phys. Lett. \textbf{B 610}, 177 (2005).

\bibitem{bib:gluonFragm} E. Braaten and T.C. Yuan,
  Phys. Rev. Lett. \textbf{71}, 1673 (1993).

\bibitem{bib:colour_evap} H. Fritzsch, Phys. Lett. \textbf{B 67}, 217
  (1977); F. Halzen, Phys. Lett. \textbf{B 69}, 105 (1977).

\bibitem{bib:gott_jack} K. Gottfried and J.D. Jackson, Nuovo
  Cim. \textbf{33}, 309 (1964).

\bibitem{bib:coll_sop} J.C. Collins and D.E. Soper,
  Phys. Rev. \textbf{D 16}, 2219 (1977).

\bibitem{bib:BrinkSatchler}
  D.M.~Brink and G.R.~Satchler, ``Angular momentum''
  (Third Edition), Clarendon Press, Oxford (1993).

\bibitem{bib:ImprovedQQbarPol} P. Faccioli, C. Louren\c{c}o and
  J. Seixas, Phys. Rev. \textbf{D 81}, 111502(R) (2010).

\bibitem{bib:LTGen} P. Faccioli, C. Louren\c{c}o and J. Seixas,
  ``Rotation-invariant relations in vector meson decays into fermion
  pairs'', arXiv:1005.2601 [hep-ph].

\bibitem{bib:NA10} S. Falciano \emph{et al.} (NA10 Coll.),
  Z. Phys. \textbf{C 31}, 513 (1986).

\bibitem{bib:E615} J.S.~Conway \emph{et al.} (E615 Coll.),
  Phys. Rev. \textbf{D 39}, 92 (1989).

\end{thebibliography}
\end{document}